\documentclass[aps,prd]{revtex4}

\usepackage{graphicx}
\usepackage{epstopdf}

\begin{document}


\pagenumbering{roman}
\thispagestyle{empty}

\begin{center}
\bf \LARGE
APS Neutrino Study:\\
\medskip
Report of the Neutrino Astrophysics and Cosmology Working Group\\
\vspace{0.2cm}
\rm \Large (29 October 2004)\\
\vspace{1cm}
\bf \LARGE Working Group Leaders:\\
\rm \Large Steve Barwick and John Beacom\\
\vspace{0.5cm}
\bf \LARGE Writing Committee:\\
\rm \Large
Steve Barwick (Irvine)\\
John Beacom (Ohio State)\\
Vince Cianciolo (Oak Ridge)\\
Scott Dodelson (Fermilab)\\
Jonathan L. Feng (Irvine)\\
George Fuller (San Diego)\\
Manoj Kaplinghat (Irvine)\\
Doug McKay (Kansas)\\
Peter M\'esz\'aros (Penn. State)\\
Anthony Mezzacappa (Oak Ridge)\\
Hitoshi Murayama (Berkeley)\\
Keith Olive (Minnesota)\\
Todor Stanev (Bartol)\\
Terry Walker (Ohio State)\\
\vspace{0.5cm}
\bf \LARGE Additional Participants:\\
\rm \Large
A.~B.~Balantekin, N.~Bell, G.~Bertone, R.~Boyd, L.~Chatterjee, M.-C.~Chen,
M.~Dragowsky, E.~Henley, A.~Karle, T.~Kattori, P.~Langacker, J.~Learned,
J.~LoSecco, C.~Lunardini, M.~Medvedev, I.~Mocioiu, P.~Nienaber,
S.~Palomares-Ruiz, S.~Pascoli, R.~Plunkett, G.~Raffelt, T.~Takeuchi,
J.~Thaler, M.~Vagins, N.~Weiner, B.-L.~Young
\end{center}


\newpage
\phantom{foobar}
\thispagestyle{empty}

\begin{figure}
\begin{center}
\includegraphics[width=7cm]{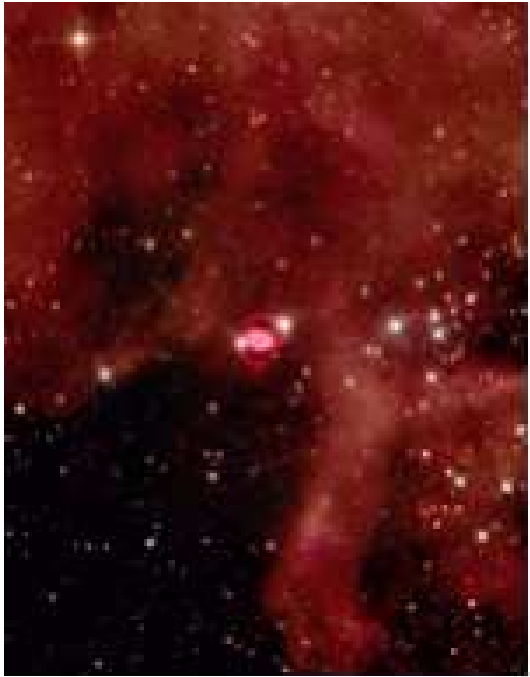}
\caption{Hubble Space Telescope image of the SN 1987A remnant in the
Large Magellanic Cloud, a close companion of the Milky Way.  Beyond the
Sun and SN 1987A, cosmic neutrino sources remain undiscovered.  An entire
Universe awaits, and the prospects for present and next-generation
experiments are excellent.}
\end{center}
\end{figure}


\newpage
\pagenumbering{arabic}
\setcounter{page}{0}

\begin{center}
{\Large\sc Introduction}\\
\end{center}

In 2002, Ray Davis and Masatoshi Koshiba were awarded the Nobel Prize
in Physics ``for pioneering contributions to astrophysics, in
particular for the detection of cosmic neutrinos.''  However, while
astronomy has undergone a revolution in understanding by synthesizing
data taken at many wavelengths, the universe has only barely been
glimpsed in neutrinos, just the Sun and the nearby SN 1987A.  An
entire universe awaits, and since neutrinos can probe astrophysical
objects at densities, energies, and distances that are otherwise
inaccessible, the results are expected to be particularly exciting.
Similarly, the revolution in quantitative cosmology has heightened the
need for very precise tests that depend on the effects of neutrinos, and
prominent among them is the search for the effects of neutrino mass,
since neutrinos are a small but known component of the dark matter.


\bigskip
{\large\bf Questions of the Neutrino Study:}\\

The Neutrino Astrophysics and Cosmology Working Group put special
emphasis on the following primary questions of the Neutrino Study; there
are strong connections to the other questions as well.

\begin{itemize}

\item {\it What is the role of neutrinos in shaping the universe?}

\item {\it Are neutrinos the key to the understanding of the
matter-antimatter asymmetry of the universe?}

\item {\it What can neutrinos disclose about the deep interiors
of astrophysical objects, and about the mysterious sources of
very high energy cosmic rays?}

\end{itemize}

The impact of neutrino physics on astrophysics and cosmology
is directly connected to many of the highest priority questions in those
fields, a case made in more detail in the next twelve sections of this report,
covering the following topics:

\begin{enumerate}

\item {\sc Origin and nature of the cosmic rays}

\item {\sc GZK neutrino detection and new physics above a TeV}

\item {\sc Neutrino probes of high energy astrophysical sources}

\item {\sc Dark matter searches using neutrinos}

\item {\sc Neutrinos as a probe of supernovae}

\item {\sc Supernova neutrinos as tests of particle physics}

\item {\sc Diffuse supernova neutrino background}

\item {\sc Measurements of neutrino-nucleus cross sections}

\item {\sc Leptogenesis and the origin of the baryon asymmetry}

\item {\sc Precision big bang nucleosynthesis tests}

\item {\sc Precision cosmic microwave background tests}

\item {\sc Neutrino mass and large scale structure}

\end{enumerate}


\bigskip
{\large\bf Working Group Recommendations:}\\

The recommendations of the Neutrino Astrophysics and Cosmology
Working Group were developed in the context of these twelve topics,
and are designed to provide strategies for answering the Questions
of the Neutrino Study identified above.  We provide recommendations
that, for a modest investment, promise very important progress in
neutrino physics, with fundamental impact in astrophysics, cosmology,
particle physics, and nuclear physics.   The recommendations focus
broadly on programmatic themes that are required to maintain the
vitality of neutrino astrophysics and cosmology while providing guidance
on the long range goals of the field.   We limited our recommendations to
those experimental concepts and/or detectors that require new money from
US funding agencies in the short term.

\begin{itemize}

\item
{\bf We strongly recommend the development of experimental techniques
that focus on the detection of astrophysical neutrinos, especially in the
energy range above $10^{15}$ eV.}

EeV ($10^{18}$ eV) neutrinos are expected from the collision of ultra-high
cosmic rays and microwave background photons. Current generation
experiments must be followed by more ambitious efforts that target specific
neutrino species or energy regimes.  In particular, we focus our
recommendation on the measurement of neutrino energies beyond
$10^{15}$ eV.  At these extreme energies, a new view of sources in
the distant, high-energy universe can be constructed; photons from
the same sources would be absorbed, and cosmic rays would be
deflected by magnetic fields.  In addition, the detection of these neutrinos
at Earth probes the energy frontier in the interaction energy, beyond the
reach of accelerators, perhaps revealing the onset of new physics.  Finally,
the sources themselves may be exotic, arising from dark matter annihilation
or decay.   In order to open this unique window on testing the particle nature
of the dark matter, astrophysical sources must be understood first.

The next generation detectors must improve on the basic optical
Cherenkov, radio Cherenkov and air shower techniques currently
employed by detectors in operation or under construction.  We also
encourage R\&D for the development of novel neutrino detection
techniques based on acoustic pulses or fluorescence flashes in order
to assess backgrounds and signal efficiency.

We estimate that the appropriate cost is less than \$10 million to
enhance radio-based technologies or develop new technologies for high
energy neutrino detection.  The technical goal of the next generation
detector should be to increase the sensitivity by factor of 10, which
may be adequate to measure the energy spectrum of the expected GZK
(Greisen-Zatsepin-Kuzmin) neutrinos, produced by the interactions of
ultra-high energy cosmic ray protons with the cosmic microwave
background.  The research and development phase for these experiments
is likely to require 3-5 years.

\begin{figure}
\begin{center}
\includegraphics[width=10cm,angle=-90]{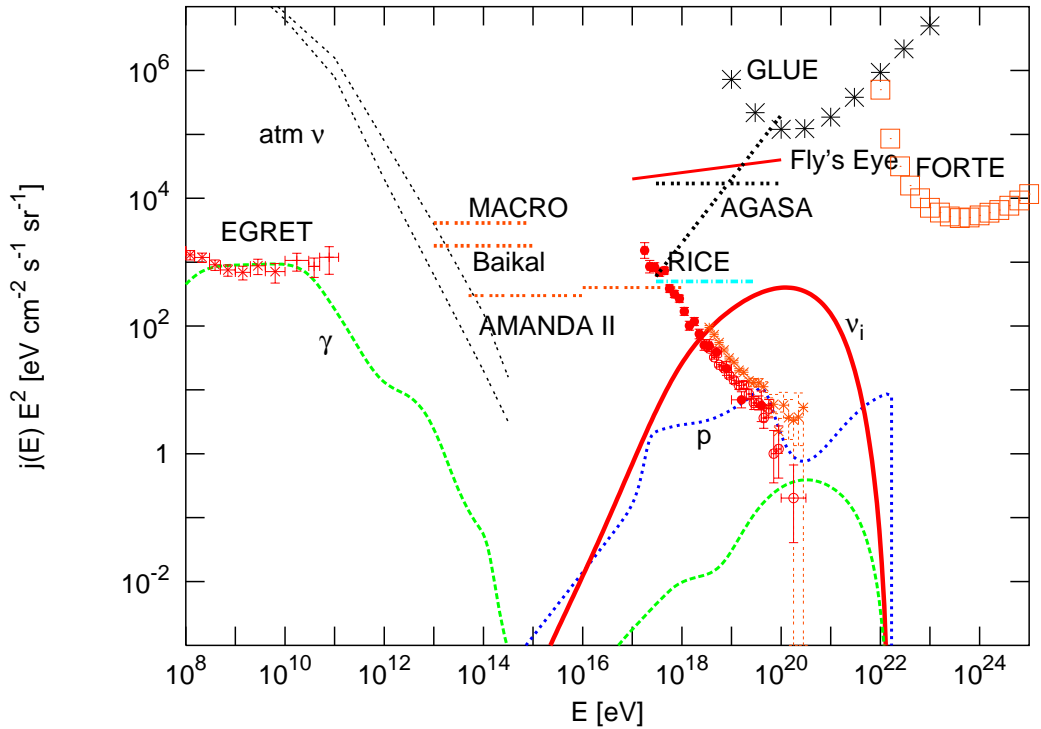}
\caption{Results are shown for the neutrino flux (solid red line)
predicted by a model of D.V. Semikoz and G. Sigl (JCAP 0404:003 (2004)
[hep-ph/0309328]), compared to existing limits (labeled by the experiments). 
This model is chosen to produce the
largest neutrino flux compatible with both the cosmic ray (red data
points, blue dotted lines) and gamma ray data (red data points, green
dashed lines), yet it remains beyond the reach of current experiments.
A new generation of experiments is is needed to test these very
important predictions, as well as to begin to survey the ultra-high
energy universe for new sources.
\label{AstroFig1}}
\end{center}
\end{figure}

\item
{\bf We recommend support for new precision measurements of
neutrino-nucleus cross sections in the energy range of a few tens of
MeV.}

These cross sections, which are uncertain at the level of several
times 10\% from theory, are very important for understanding
supernovae, specifically the neutrino opacities, nucleosynthesis, and
detection.  For example, in the event of a Milky Way supernova,
Super-Kamiokande would observe $\sim 10^3$ events from neutrino
scattering on oxygen nuclei.  Those cross sections, which have never
been measured, are crucial for the interpretation of the data, and
especially for the effects of neutrino mixing.  These measurements
would be of fundamental interest for constraining nuclear models, and
measurements on a few different nuclei would reduce theoretical
uncertainties for the cross sections on unmeasured nuclei.  Low-energy
neutrino-nucleus experiments would also provide important constraints
on cross sections in the GeV range, important for future sensitive
oscillation studies at accelerators, for example regarding visible
de-excitations of the residual nucleus.  The availability of an
intense source of neutrinos, with energy spectra matching those from
supernovae, combined with the strong current interest in neutrinos for
supernova science, makes a compelling case for the development of a
neutrino-nucleus scattering research facility.  At present, the only 
realistic near-term possibility would be the stopped-muon source at the
Spallation Neutrino Source.  Ultimately, it may also be possible to
create a low-energy beta-beam source, and this possibility, and its
connection to a possible Rare Isotope Accelerator, should be
monitored in the future.  Ideally, this Neutrino Study will consider a 
comprehensive approach to the issue of measuring neutrino cross sections, 
both for their intrinsic and practical interest.

We estimate that measurements of neutrino cross-section recommended by
this working group can be accomplished for less than \$10 million,
with R\&D requiring \$0.5 million for one year.  Construction will
require two additional years.

\item 
{\bf We recommend that adequate resources be provided to allow
existing large-volume solar, reactor, proton decay, and high energy
neutrino telescopes to observe neutrinos from the next supernova
explosion and participate in a worldwide monitoring
system. Furthermore, future large-volume detectors should consider the
detection of supernova neutrinos an important science goal and plan
accordingly.}

Core-collapse supernovae are prodigious sources of neutrinos with
energies of a few tens of MeV.    Though only about 20 neutrinos were
observed from Supernova 1987A, those precious few events have led
to literally thousands of published papers.   The observation of neutrinos
from another supernova is one of the most important goals of particle
and nuclear astrophysics.  With much larger detectors, and much higher
statistics expected, the scientific payoff would be significantly greater.   Since
supernovae in the Milky Way are rare, at a rate of a few per century,
it is vitally important that existing and future detectors, built for other
purposes, be able to observe a neutrino burst with maximum livetime
and detector efficiency; several detectors are needed, to maximize
the certainty of detection, and for their complementary abilities.  It is
also extremely important to detect the diffuse supernova neutrino
background, the fossil record of all of the supernovae in the universe.
The limit from Super-Kamiokande is just above theoretical predictions
that are normalized to the measured star formation rate as a function 
of redshift.

We anticipate that the investment to insure that large volume
detectors maintain sensitivity to galactic supernovae, as well as the
diffuse supernova neutrino background from all supernovae, will be
less than \$10 million over the next 5 years.  New large volume
detectors expected for long-baseline, reactor, proton-decay, solar,
and high energy neutrino detectors should consider new ideas to
enhance the capabilities for the detection of supernova neutrinos.
The cost is not possible to determine at this time.

\end{itemize}


\bigskip
{\large\bf Working Group Endorsements:}\\

In addition to our recommendations, we wish to express our strong and
enthusiastic endorsement of four other science goals in neutrino physics.

\begin{itemize}

\item
{\bf We enthusiastically support continued investment in a
vigorous and multi-faceted effort to precisely (but indirectly)
measure the cosmological {\em neutrino} background through its effects
on big-bang nucleosynthesis, the cosmic microwave background, and the
large-scale structure of galaxies; in particular, weak gravitational
lensing techniques offer a very realistic and exciting possibility of
measuring neutrino masses down to the scale indicated by neutrino
oscillations.}

Big bang nucleosynthesis (BBN) and cosmic microwave background (CMB)
observations are each sensitive to the number of neutrino flavors, and the
present constraints are roughly $N_{BBN} = 2-4$ and $N_{CMB} = 1-6$,
respectively, in  agreement with accelerator data.   Next-generation
observations of the primordial abundances of deuterium, helium and
lithium will improve the precision of BBN, testing both the standard model of
particle physics and the framework of standard cosmology.   Though less
sensitive now, the CMB constraint on the number of neutrino flavors is
expected  to markedly improve, to an uncertainty of well less than 1 equivalent
neutrino.   Note that extra particles, such as sterile neutrinos, could add about 1
additional flavor, so that these measurements are extremely important for
testing whether the three-flavor oscillation picture is complete.

Neutrinos are the only known component of the non-baryonic dark matter.
The present cosmological limit on neutrino mass, coupled with the measured
mass-squared differences from solar and atmospheric neutrino data, is
presently at the level of $0.3-0.6$ eV, a few times more stringent than limits
inferred from the tritium beta-decay experiments.  Future cosmological tests
of neutrino mass with galaxies and the cosmic microwave background have
excellent prospects for reaching the mass scale of $\sqrt{dm^2_{atm}} \sim 0.05$
eV, by which the discovery of at least one neutrino mass is guaranteed. 
Cosmological and astrophysical data provide a novel suite of tools to determine
neutrino properties and simultaneously provide an independent cross-check
for laboratory tritium beta decay and neutrinoless double beta experiments.
If neutrino mass is discovered by cosmological observations, it will confirm 
our assumption that the relativistic particle background required by BBN and
the CMB is indeed composed of neutrinos.

Leptogenesis models connect neutrinos to the unexplained matter
dominance of the universe, and may also connect light neutrino masses
to GUT-scale physics.   A crucial observable for these models is the scale
of the light neutrino masses.

\item
{\bf We enthusiastically support theoretical and computational
efforts that integrate the latest results in astronomy, astrophysics,
cosmology, particle physics, and nuclear physics to constrain the
properties of neutrinos and elucidate their role in the universe.}

Theory plays an especially important role in integrating and interpreting the
progress in all the fields represented by this working group.  To fully realize the
benefits accrued from these ever-growing connections, it is essential
to adequately fund the theoretical community.

It has become increasingly clear that important astrophysical phenomena,
such as core collapse supernovae, in which neutrinos play a central role 
and which can be used to probe the properties of neutrinos under conditions
not accessible in terrestrial experiment, are multiphysics phenomena that require
large-scale, multidisciplinary collaboration and computation for their
understanding.  This presents a new paradigm for theoretical investigation, 
resembling more the longer-term, larger-scale investigations traditionally
supported under experimental science programs than the single-investigator,
or small-group-investigator, efforts traditionally support by theory programs.  
This new investigation paradigm presents, therefore, a new dimension to
the process of setting priorities for future investigations in neutrino physics
and astrophysics, and should be taken into account.

\item
{\bf We enthusiastically support the scientific goals of the
current program in galactic and extra-galactic neutrino astrophysics
experiments, including Super-Kamiokande, AMANDA, and NT-200 deployed
in Lake Baikal.  Furthermore, we endorse the timely completion of
projects under construction, such as IceCube, undersea programs in the
Mediterranean, ANITA, and AUGER.}

\item 
{\bf Though solar neutrinos were not in our purview, we endorse
the conclusion of the Solar/Atmospheric Working Group that it is
important to precisely measure solar neutrinos, and strongly support
the development of techniques which could also be used for direct
dark matter detection.}

\end{itemize}


\bigskip
{\large\bf Preface to the Twelve Working Group Topics:}\\

Neutrino detection is a tough business.  When first proposed as a
fundamental constituent of matter back in the first half of the
twentieth century, the brightest minds in experimental physics
generally considered the neutrino impossible to observe.  Wolfgang
Pauli, who first postulated the neutrino, rued his creation, because
he though that he had invented a particle that could not be
discovered.  When Frederick Reines and Clyde Cowen proved Pauli wrong
on this account, the detection of the neutrino was considered so
significant and such an experimental tour-de-force that a Nobel Prize
was awarded in 1995. Why go through such heroic efforts to detect and
measure the properties of the neutrino?  This report hopes to answer
this question from the perspective of astrophysics and cosmology.
 
Yogi Berra once quipped, ``You can see a lot just by looking", and the
neutrino presents us with a powerful tool to look deep into the heart
of the most explosive objects in the cosmos and deep into the far
reaches of the universe.  The importance of the neutrino in astrophysics
was quickly recognized. In 1938, Bethe and Critchfield outlined a
series of nuclear reactions deep in the interior that provided the
first realistic mechanism to power the sun.  Unfortunately, it is not
possible to use conventional optical telescopes to look into the
center of the sun.  However, detailed calculations by John Bahcall in
the 1960's led to the idea that the neutrino, a copious byproduct of
the nuclear reactions, could be detected and used to directly verify the
theory of solar energy.  The experimental detection of solar neutrinos
by Ray Davis and the subsequent development of solar neutrino
astronomy by Masatoshi Koshiba resulted in yet another Nobel Prize in
2002. These pioneering efforts first uncovered an interesting
conundrum: the detection rate of solar neutrinos was less than half of
the expected rate in water detectors and only a third in the Chlorine
detector of Ray Davis. Was this a problem with our understanding of
the sun or was there some new fundamental physics about neutrinos that
had not been uncovered by terrestrial experiments? We now know that
these early experiments provided the first hint that neutrinos have
non-zero mass and the morphing between the electron neutrino and other
types was enhanced by the matter of the sun, an effect referred to the 
Mikheyev-Smirnov-Wolfenstein (MSW) effect in honor of its discoverers.  The 
exciting history of the detection and impact of
solar neutrinos is the subject of the report by the Solar and
Atmospheric Neutrinos Working Group.

While the story of solar neutrinos provides the first example of the
intimate connection between astrophysics and fundamental particle
physics, the discussion in the subsequent sections of this report
demonstrate that it is far from unique.  Today, astrophysicists widely 
believe
that neutrinos will play an important role in deciphering the energy
sources that drive the most powerful objects in our galaxy and beyond,
such as supernovae, black holes, Active Galactic Nuclei, and
Gamma Ray Bursts. Astrophysicists hope that high energy neutrinos
point back to the sources of cosmic rays, which are enormously
energetic particles that rain down on earth.  A few cosmic rays are
observed to possess energies that are nearly a million times LARGER
than produced by Fermilab or LHC at CERN.  But how are they made and
where do they come from?  We do not know at present and the mystery is
only deepening.  Detection of neutrinos from astrophysical sources
would provide insight on the longstanding question of the origin of
highest energy cosmic rays.  The pioneering AMANDA high energy
neutrino telescope, located more than a mile beneath the snow surface
at the South Pole, is designed to search for astrophysical sources of
neutrinos at TeV energies.  Both AMANDA, and its successor, IceCube,
are beautiful examples of productive international collaboration.
Several scientific panels have discussed the scientific potential of
high energy neutrino astronomy, including the report by the NRC
Astronomy and Astrophysics Survey Committee and the NRC report
``Neutrinos and Beyond".  We should also point out that several
(predominantly) European efforts to construct a neutrino telescope in
the Mediterranean Sea are currently underway.

The rich potential of multi-messenger astronomy remains an enticing
promise even though no extrasolar sources of neutrinos have been
detected, except for the supernova that was observed in 1987. There is
intriguing indirect evidence which suggests that existing experiments
like AMANDA, or those now under construction, are tantalizingly close
to the sensitivity required to detect astrophysical neutrinos.
Collectively, current generation and approved detectors observe
neutrino energies that span over 10 orders of magnitude.

Although the GZK mechanism provides a compelling theoretical prediction 
for extragalactic neutrinos, there may well be other important sources of 
extremely high energy neutrinos.  For example, one idea discussed in the 
literature involves the decay of supermassive particles.  This and other 
exotic ideas for neutrino production highlight the richness of the physics 
potential of astrophysical and cosmological probes.  Furthermore, if any 
source can produce neutrino energies that extend up to $\sim 10^{22}$ eV, 
then it becomes possible to directly observe the cosmological neutrino
background, a  residue from the Big Bang, by detecting an absorption
feature at the Z-boson resonance.

Arguably the most exciting developments in physics during the past
decade evolved from the study of distant supernovae and the cosmic
microwave background.  From these studies, we now know that the
universe contains a surprising mixture of ordinary matter, dark matter
of unknown identity, and dark energy of unknown physics.  We also know
that the Universe contains far more matter than anti-matter, a
situation that is not obvious from the interactions observed in
earthbound accelerator experiments. If one assumes that the early
universe was symmetric with respect to matter and anti-matter, then at
one point, the preponderance of matter over anti-matter must be
created dynamically.  In 1967, Sakharov pointed out that one possible
mechanism required the violation of the CP symmetry and baryon number.
CP violation was indeed discovered by Cronin and Fitch, but searches
for the violation of baryon number have all failed.  In particular,
the violation of baryon number is required for protons to decay into
less massive particles, but the proton remains stubbornly stable.  An
alternative proposal is known as leptogenesis.  Here it is assumed
that early in the history of the universe, a preponderance of leptons
(e.g., electrons and neutrinos) over anti-leptons was produced.  It is
then possible within the standard model to transfer the excess in
leptons to an excess in all matter by non-perturbative effects that
conserve the difference between the net baryon and lepton number, 
$B-L$.  For this scenario to work there must be lepton number violation and CP
violation specifically associated with leptons.  The quest to measure
the degree of CP violation in lepton interactions is a major physics
goal of the neutrino community.

Big Bang cosmology predicts that neutrinos outnumber protons and
nuclei by about a billion.  Since we also know from recent studies of
the atmospheric and solar neutrino fluxes that neutrinos have mass,
the residual neutrinos from Big Bang constitute part of the dark
matter of the Universe, the only dark matter constituent identified so
far. Present limits on neutrino mass tell us that neutrinos are only a
small part of the dark matter, but even relatively small masses can
influence the structure and patterns of clustering of galaxies and
fluctuations of the cosmic microwave background.

It is really no surprise that the electrically neutral and nearly
massless neutrino provides an astonishing breadth of opportunities for
astrophysicists and cosmologists.  Neutrinos interact solely by the
weak interaction, the only known stable particle with this property.
Consequently, astrophysicists can detect neutrinos that begin their
journey from any point in the Universe.  Once produced, they can
escape from the hot dense cores of Active Galactic Nuclei or exploding
supernova, and then travel to earth unimpeded by anything else.
They are not deflected by magnetic fields, so they travel in
straight lines.  The messages they carry are key to understanding the
internal engines that drive these distant beacons. It is safe to say
that a more complete understanding of all the fundamental physics
properties of the neutrino provides the greatest chance to extract
as much as possible from the neutrino messenger.
 
Neutrino properties such as neutrino mass, oscillation, and perhaps
most importantly, unanticipated interaction mechanisms can be probed
over a broad range of environmental conditions found throughout the
cosmos. Experiments that utilize astrophysical neutrinos can survey
large patches of parameter space, and help to provide insight on where
to focus the next generation of terrestrial experiments. Neutrino mass
impacts subtle details of fluctuations in the cosmic microwave
background radiation, and neutrino oscillation implies that cosmic
accelerators will illuminate the earth with beam containing all
neutrino flavors.  This report summarizes several, but by no means
all, of the important ideas and experimental techniques that are
poised to take advantage of the opportunities that nature
provides. Moreover, we highlight the complementary role of neutrinos
from astrophysical phenomena in leading to breakthroughs in the
understanding of neutrino properties and the measurement of neutrino
properties by accelerator and reactor experiments in the
interpretation of astrophysical data.

The following twelve sections explore these rich topics in neutrino
astrophysics and cosmology in more detail.


\newpage
\begin{center}
{\Large\sc Origin and nature of the cosmic rays}\\
\end{center}


 The assumption that the Ultra High Energy Cosmic Rays (UHECR) are
 nuclei (presumably protons) accelerated in luminous extragalactic
 sources provides a natural connection between these particles and
 ultra high neutrinos. This was first realized by
 Berezinsky\&Zatsepin~\cite{BerZat69} soon after the introduction of
 the Greisen-Kuzmin-Zatsepin (GZK)
 effect~\cite{GZK}. The first  realistic calculation of the
 generated neutrino flux was made by  Stecker~\cite{Stecker73}.
 The problem has been revisited many times after the paper of
 Hill\&Schramm~\cite{HS85} who used the non-detection of such neutrinos
 to limit the cosmological evolution of the sources of UHECR. 
 
 Cosmological neutrinos are produced in interactions of the UHECR with
 the ambient photon fields, mostly with the microwave background
 radiation. The GZK effect is the limit on the high energy extension 
 of the cosmic ray spectrum in case their sources are isotropically
 and homogeneously distributed in the Universe. The physics of these
 photoproduction interactions is very well known. Although the energy
 of the interacting protons is very high, the center of mass energy
 is low, mostly at the photoproduction threshold. The interaction
 cross section is studied at accelerators and is very well known.
 Most of the interactions happen at the $\Delta^+$ resonance where
 the cross section reaches 500$\mu$b. The mean free path reaches a
 minimum of 3.4 Megaparsecs (Mpc) at energy of 6$\times$10$^{20}$ eV.
 The average energy loss of 10$^{20}$ protons is about 20\% per
 interaction and slowly increases with the proton (and center of mass)
 energy.

 The fluxes of cosmological neutrinos are, however, very uncertain
 because of the lack of certainty in the astrophysical input.
 The main parameters that define the magnitude and the spectral shape
 of the cosmological neutrino fluxes are: the total UHECR source
 luminosity $L_{CR}$, the shape of the UHECR injection spectrum
 $\alpha_{CR}$, the maximum UHECR energy at acceleration $E_{max}$
 and the cosmological evolution of the UHECR sources. These are the
 same parameters that Waxman\&Bahcall~\cite{WB1} used to set a limit
 on the neutrino fluxes generated in optically thin sources of UHECR.
 We will first use the parameters of this limit to compare the
 cosmological to source neutrinos.

\begin{figure}
\includegraphics[width=6truein]{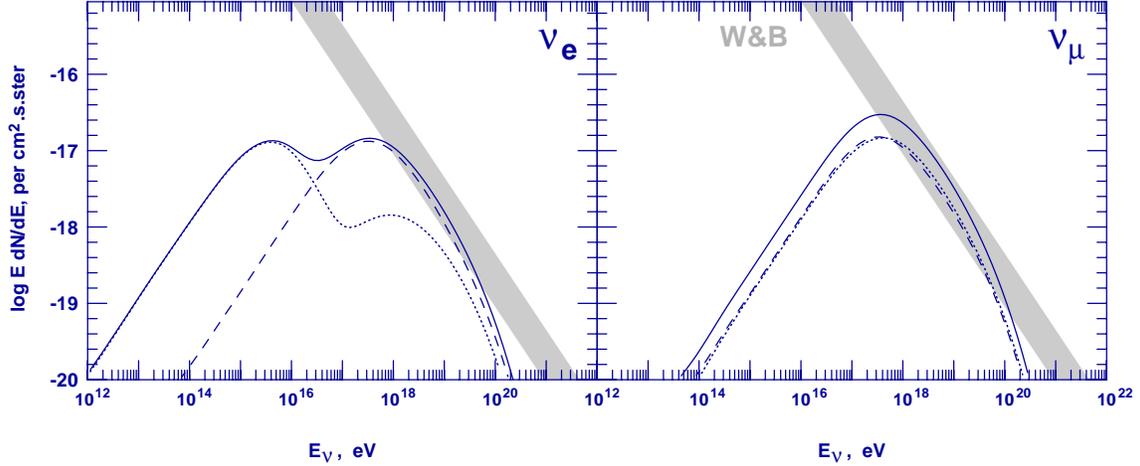}
\caption{The flux of cosmological neutrinos is compared to the
 Waxman\&Bahcall limit for source \protect$\nu_\mu + \bar{\nu}_\mu$,
 which is plotted as a shaded band. The upper edge of the band is
 for source luminosity evolution as \protect$(1+z)^3$ and the lower
 one is for no cosmological evolution. 
\label{standard}}
\end{figure}

 Waxman\&Bahcall use cosmic ray source luminosity $L_{CR} =
 4.5 \pm 1.5 \times {\rm 10}^{44}$ erg/Mpc$^3$/yr between
 10$^{19}$ and 10$^{21}$ eV for power law  with $\alpha$ = 2.
 The assumption is that no cosmic rays are accelerated above
 10$^{21}$ eV. The cosmological evolution of the source luminosity
 is assumed to be $(1+z)^3$ to $z$ = 1.9 then flat to $z$=2.7  with
 an exponential decay at larger redshifts.
 Fig.~\ref{standard} shows the cosmological neutrino
 fluxes that correspond to these input parameters~\cite{ESS01}.
 Cosmological model used is with $\Omega_\Lambda$ = 0.7 and
 $\Omega_M$ = 0.3 and $H_0$ = 75 km/s/Mpc.

 At energy about 3$\times$10$^{18}$ eV the cosmological fluxes
 of $\nu_\mu + \bar{\nu}_\mu$ are very close to the limit for
 source neutrinos. The reason is simple - in propagation from 
 large distances protons lose almost all of their energy in
 interactions on MBR. An interesting feature is the flux of 
 $\bar{\nu}_e$ which peaks at energy about 3$\times$10$^{15}$ eV.
 The origin of this flux is neutron decay, and a small 
 $\bar{\nu}_e$ flux is generated in neutron interactions on MBR.
 
 The cosmological evolution of the sources (n=3) increases the 
 fluxes by about a factor of five. The increase, however, is
 energy dependent.  The highest energy neutrinos are generated 
 at very small redshifts. The low energy neutrinos come from
 high redshifts because of two reasons: the threshold energy of
 protons for photoproduction interaction decreases, and the
 generated neutrinos are further redshifted to the current epoch
 This flux would generate about 0.4 neutrino induced showers 
 per year in the IceCube~\cite{IceCube} neutrino detector 
 and 0.9 events in the Auger\cite{Auger} observatory (for target mass
 of 30 km$^3$ of water) assuming that at  arrival at Earth the
 flavor ratio $\nu_e : \nu_\mu : \nu_\tau$  is 1:1:1 because of
 neutrino oscillations. 
 ANITA is expected to observe several events per year.
 It is difficult to estimate the rate in
 EUSO~\cite{EUSO} because of its yet unknown energy threshold.
 These events come
 from the NC interactions of all neutrinos, CC interactions of
 $\nu_e$, the hadronic ($y$) part of the CC interactions of muon
 and tau neutrinos and from $\tau$ decay. The Glashow resonance 
 does not produce high rate of events because of its narrow 
 width. Ice Cube should also detect very energetic muons with
 a comparable rate which is difficult to predict in a simple way.
 
The expected flux of GZK neutrinos depends on several ingredients:

\begin{itemize}
\item{\em UHECR luminosity}
 The source luminosity derived by Waxman is, however, very uncertain.
 It is reasonable when the normalization is at 10$^{19}$ eV, but it
 would easily increase by a factor of three
 if the majority of the cosmic rays above 3$\times$10$^{18}$ are also
 of extragalactic origin, On the other hand, if the local flux 
 of UHECR is higher than the average in the Universe (since the matter
 density in the local Universe is somewhat higher than the average one)
 the luminosity could easily decrease by a factor of two.

\item {\em Maximum injection energy}
 The calculation shown in Fig.~\ref{standard} is done with $\alpha$=2
 power law spectrum extending to 10$^{22}$ eV with exponential cutoff
 at 10$^{21.5}$.  A decrease of the maximum acceleration energy by 
 a factor of 10 would decrease significantly the cosmological
 neutrino flux as the number of interaction proton decreases.

\item {\em Injection spectrum}
 Most of the analyses of the injection spectrum that generates 
 the observed UHECR after propagation estimate an injection spectrum 
 not flatter than a power law with $\alpha$ = 2.5. The extreme case
 is developed by Berezinsky et al~\cite{BerGazGri02} who derive 
 an $\alpha$=2.7 injection spectrum. The luminosity required for
 the explanation of the observed events in the 10$^{19}$ - 10$^{20}$
 range then becomes 4.5$\times$10$^{47}$ erg Mpc$^{-3}$yr$^{-1}$.
 A steeper spectrum would generate only a small event rate for the
 giant air shower arrays.

\item {\em Cosmological evolution}
  Expressed in terms of $(1 + z)^n$ the cosmological evolution of
 different objects is observed to be between $n$=3 and 4. 
 A strong evolution with $n$ = 4 will not only increase the
 total flux of cosmological neutrinos, but will also move the 
 maximum flux to somewhat lower neutrino energy, since the 
 contribution at large redshift increases.
 
\item {\em Other photon targets}
   Finally the MBR is not the only universal photon target.
 Especially interesting is the isotropic infrared and optical
 background (IRB). Its total density is significantly smaller
 than that of MBR. Recent models of IRB give 1.6 photons/cm$^3$,
 a factor of 250 less than the MBR.  On the other hand, 
 protons of lower energy can interact on the IRB, and the
 smaller number density has to be weighted with the larger flux
 of interacting protons. The present Universe is optically thin
 to 10$^{19}$ eV and lower energy protons, but even at small
 redshift the proton interaction rate quickly increases. 
 The probability of 10$^{18}$ eV proton interactions could also be
 increased if the UHECR sources are in regions of high magnetic field
 and infrared background density.
\end{itemize} 

  The estimated shower event rates above 10$^{15}$ eV  per km$^3$yr
 vary from 0.2 and 1.2 for the Waxman\&Bahcall luminosity function.
 The lowest rate corresponds to local UHECR density exceeding
 the average in the Universe by a factor of two, flat power law
 injection spectrum ($\alpha$=2), and $(1+z)^n$ cosmological evolution.
 This rates corresponds to one half of the fluxes shown in
 Fig.~\ref{standard}.
 The higher rate is achieved by assuming that we see the average
 UHECR density, the injection spectrum is with $\alpha$=2.5, and
 the cosmological evolution is with $n$=4. It also includes
 interactions on the infrared background radiation. The corresponding
 event rates for shower energy above 10$^{19}$ eV, which are suitable
 for the Auger observatory vary between 0.44 and 0.66 for 30 km$^3$ 
 of water target. Both event rates would increase by approximately
 the same coefficient if the UHECR luminosity were higher. 
 The difference between the event rates reflects the shape of the
 UHECR injection spectrum and could be further affected by an increase 
 or decrease of the maximum injection energy. 

 The possible detection of cosmological neutrinos should be 
 analyzed together with the shape of the UHECR spectrum above
 the GZK cutoff at 4$\times$10$^{19}$ eV.  Here the AUGER results
 are eagerly awaited.  If it confirms with 
 high experimental statistics the GZK feature, as claimed by
 the HiRes experiment~\cite{HiRes}, the flux of cosmological
 neutrinos would reveal the distribution of sources that generate
 UHECR in our cosmological neighborhood. If there is no observed 
 cutoff (as claimed by the AGASA~\cite{AGASA} experiment), the 
 flux of cosmological neutrinos would be much smaller and possibly
 undetectable. This would be an argument in favor of the top-down
 exotic scenarios for the UHECR origin. Such scenarios inevitably
 predict specific UHE neutrino spectra.
 
  In the most optimistic, although not unrealistic, case that UHECR
 sources are embedded in regions of high magnetic field and 
 ambient photon density, the detection of even single UHE neutrino
 could help reveal the source direction.


\newpage
\begin{center}
{\Large\sc GZK neutrino detection and new physics above a TeV}\\
\end{center}


{\em The GZK cutoff}: Four decades ago, two completely unexpected and
unrelated discoveries - ultra high energy (UHE) cosmic rays and the
cosmic microwave background radiation (CMB) combined to open new
windows on the Universe.  Within several years of the CMB discovery,
Greisen and, independently, Zatsepin and Kuzmin (GZK) pointed out
that UHE cosmic ray protons interacting with the CMB above photo-pion
production threshold would lose energy until they fell below
threshold.  They found that experiments should see an upper cutoff in
the spectrum in the interval $10^7$ TeV to $10^8$ TeV, or $\sqrt{s}
\approx$ 150 TeV to 450 TeV, if the sources lay beyond 50 Megaparsecs
or so.

{\em The GZK neutrinos}: Past and ongoing cosmic ray experiments agree
that there are tens of events above $10^7$ TeV with fluxes of roughly
one per square kilometer per century.  Experiments disagree whether
enough events sufficiently above $10^8$ TeV have been seen to indicate
violation of the cutoff \cite{agHi}.  The status of the GZK cutoff is
an open question, which may be answered by the AUGER experiment within
the next several years.  Regardless of the resolution to the GZK
cutoff question, the mere fact that the UHE cosmic rays exist has a
crucial impact on the prospects for opening up UHE neutrino
astrophysics at neutrino energies above $10^5$ TeV.  (This
translates to 15 TeV in $\sqrt{s}$ - the LHC with hadron+lepton! The
highest laboratory energy charged current cross section measurement at
$\sqrt{s}$ = 320 GeV from DESY corresponds to $E_{\nu}$ = 50 TeV.)
  
\begin{figure}[t]
\includegraphics[height=12cm]{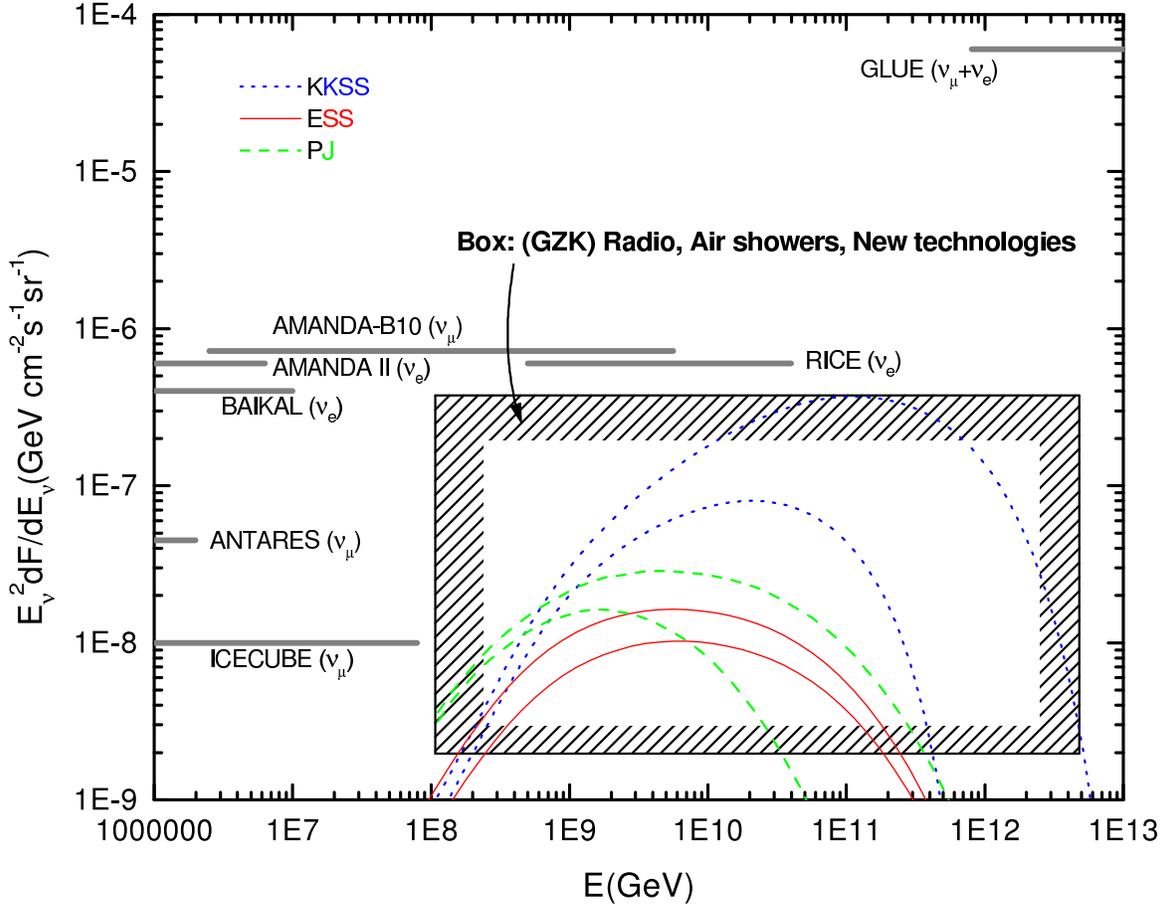}
\caption{GZK $\nu_{e}$ flux models from Protheroe and Johnson, Kalashev
et al. and Engel, Seckel, and Stanev \cite{fluxes}, covering a wide
range of values, are shown with current limits (AMANDA, RICE,
Baikal, and GLUE) and projected flux sensitivities  (ANTARES and ICECUBE)
and corresponding energy ranges.   We chose references that
present limits for $E^{-2}$ effective fluxes over a given energy range.
It should be kept in mind that these rough sensitivity projections depend
on uncertain assumptions and varying conventions.
\label{limits}}
\end{figure}

Accepting the GZK picture, one anticipates a ``guaranteed'' flux of
UHE neutrinos, as presented in Stanev's cosmological flux section in
this working group's report. Though the general arguments for the
existence of the GZK flux are more than reasonable, the shape and
normalization of the spectrum is not nailed down.  Currently available
models allow a wide range of spectra that are consistent with
constraints from photon fluxes and from the cosmic ray data
itself. Clearly the first goal is to establish the existence of a UHE
neutrino flux and a number of first surveys have already established
limits on fluxes from $10^3$ TeV into and above the GZK region
\cite{limit}. The good news is that {\em the second generation of
existing detectors and new detectors in the design and development
stage promise to cover most of the flux estimates}.  The continuation
of this expansion of the ``cast a broad net'' effort is completely
justified in light of advances so far and future projections
\cite{projections}, as illustrated in Fig. \ref{limits}. The challenge
for the future is to find ways to fight the rapidly falling fluxes
with increasing effective areas/volumes to reach observable levels as
energy rises.
                                                    
{\em UHE neutrino world}: The first detection of UHE neutrinos will be
an electrifying event in its own right and profound in its implication
for explorations of cosmologically distant sources and neutrino
interactions.  It is precisely because the interactions of neutrinos
are weak at low $\sqrt{s}$ that they can reveal information from their
sources directly, arriving along the line of sight.  For the same
reason, they are prime candidates to reveal new physics at high
$\sqrt{s}$ when they collide with nucleon targets in detectors.  The
contrast between expected and unexpected physics is not buried under
strong interactions

Post discovery possibilities include studies of the correlation of
neutrino events with known high energy sources like quasars or gamma
ray bursts. Pointing with error cones at the few degree level or
better, typical of highest energy cosmic ray events, should be
sufficient to find or reject correlation (or, more conservatively,
find or reject absence of correlation - a less model dependent
hypothesis). Comparing the neutrino event characteristics, energy
spectrum and directions with the cosmic ray shower data should reveal
much about the sources and about the identity, whether standard or
not, of the cosmic rays.  The proton hypothesis for the high end of
the spectrum is consistent with current data, but a unique
identification is needed.

The difficult job of disentangling neutrino physics from the neutrino
fluxes in data analysis will begin in earnest after discovery,
with design and deployment of enhanced second generation or third
generation detectors. In the course of this evolution, the most
effective, complementary techniques should emerge, along with a better
understanding of backgrounds, and the elimination of models giving
overly optimistic flux estimates.  One expects that advances in
neutrino physics will be accelerated by advances in cosmic ray physics
in general.  For example the constraints on the GZK flux possibilities
will increase as the gamma and the hadronic cosmic ray data improve.

Angular distributions of events can go far to resolve the interactions
from the fluxes. This demands a significant number of events with
reasonable angular resolution, since binning in angle near the horizon
will be required for the events in the GZK range.  At energies in the
$10^3$ TeV to $10^4$ TeV range, fluxes from the low end \cite{wb} to
the high end \cite{proth} of currently available estimates, would
allow a $KM^3$ detector or ANITA to discriminate the standard model cross
sections from those resulting from a rapid rise starting at the 1- 2
TeV scale. An attractive theoretical proposal with this feature is low
scale gravity \cite{lsg}, recognized as potentially significant for
UHE neutrino astrophysics \cite{d+d}. It was further realized that
production of micro black holes should set in when $\sqrt{s}$
exceeds the extra-dimensional gravity scale, potentially increasing
the $\nu - N$ cross section by a large factor \cite{f+s}.

GZK neutrino fluxes do not extend down to the $10^3$ TeV to $10^4$ TeV
regime, the top of the range where there is still roughly 30 degrees
below the horizon where useful numbers of neutrinos may be detectable.
In the GZK regime, rather good angular resolution combined with much
larger effective detector volumes would be needed to do the job, which
may be achievable with the air shower observatories or large volume
surveys in ice, salt or water in combination with established
techniques extended upward from lower energy to provide calibration.
With downward event rates alone, establishing the spectrum and using
general shape features expected of cosmogenic neutrino fluxes, cross
sections that rise radically faster than the standard model cross
sections might be distinguished. This question is especially
interesting for low scale gravity with 5 or 6 large extra dimensions,
which are relatively unconstrained by other astrophysical and
accelerator data.  Several groups have pursued this question and find
that current detection limits set constraints on a range of scenarios
for neutrino fluxes and black hole physics \cite{olinto}.

{\em Other new physics connections}: The new physics emphasis above
has been on the GZK flux and the new physics of neutrinos that may be
discovered by their interactions. Entirely different questions can
also be addressed. Bounds on TeV-scale WIMPS annihilating in the
center of the Earth have been set by searching for non-atmospheric,
upcoming neutrinos at AMANDA \cite{wimps}.  The extension of this
``TeV - physics'' search is an important tie-in with dark matter
physics for the extended programs in the $E_{\nu}$ = 1-100 TeV
range. Detectors can also be ``multi-taskers'', looking for monopoles,
which have distinctive signatures in certain detectors, or post GZK
neutrinos that may originate in a ``top down'' picture like
topological defects with masses of order, $10^{14}$ GeV, and
``hidden'' from hadrons and photons.  Extreme demands on detectors
would be required to do physics with sources at these energies;
proposals include search for direct evidence of the cosmic neutrino
background, determination of the absolute value of neutrino masses
\cite{ersw}, and distinguishing between Majorana and ``pseudo-Dirac''
neutrinos at the $10^{-18}$ eV level \cite{pseudo}! Among other
fundamental questions that may be accessible with UHE neutrino
experiments is the limit to which Lorentz symmetry violation can be
pushed \cite{cg}.

This is the high energy neutrino frontier field.  The experimental
efforts are already numerous and varied.  Continued support for
development and deployment of current techniques and research must be
continued if the breakthrough to first observations is to be made.
Serious investment in exploratory methods like acoustic, radio, and
offshoots will be needed to achieve future data sets big enough to do
detailed science.


\newpage
\begin{center}
{\Large\sc Neutrino probes of high energy astrophysical sources}\\
\end{center}

\def\Mesz{M\'esz\'aros}


Active galactic nuclei (AGN), gamma-ray bursts (GRB), and related
objects such as supernovae (SN), black holes (BH) and neutron stars
(NS) are among the most energetic sources in the Universe, involving
energy densities and gravitational fields far surpassing anything
achievable in the laboratory. Yet the phenomenology and theoretical
understanding of these high-energy sources has been severely limited
by the fact that our information about them has been obtained almost 
exclusively through the electromagnetic channel. This extends in some 
cases up to tens of GeV, and in rare cases to TeV energies \cite{weekes04}.
However, these sources are thought to be also copious emittors of 
gravitational waves, as well as of cosmic rays and neutrinos, whose 
energy fluxes may rival that in the electromagnetic channel. 
Furthermore, in the latter two cases the particle energies may reach 
\cite{w95,rb93} up to $\sim 10^{20}$ eV (the GZK range), which 
exceeds by up to eight orders of magnitude that of the most energetic 
photons detectable.  The amount of information available in these new 
channels is of a completely different nature than that so far available. 

Besides providing tests of fundamental physics extending up to the
so far unexplored PeV center-of-momentum energy range, ultra-high 
energy neutrino measurements could yield crucial insights into the 
origin and propagation of cosmic rays, and would provide a unique 
probe into the nature of these high energy astrophysical sources.  
They would directly probe both the hadronic content of the jets inferred 
in AGN and GRB, and the cosmic ray acceleration process thought to give 
rise to the diffuse cosmic ray background, the atmospheric neutrino
background, and a portion of the MeV to multi-GeV gamma-ray background. 
Neutrinos in the TeV-EeV range would mainly arise from photomeson 
interactions between protons and intra-source photons, created either 
from synchrotron and inverse Compton processes or from hadronic decay 
cascades followed by the same processes.  In the former case, the photons 
are linked to electrons or positrons thought to be accelerated in shocks, 
or possibly magnetic reconnection sheets. The same accelerators would 
unavoidably also accelerate protons, if these are present in the same
regions. In addition, $\lesssim$ GeV neutrinos may also be produced in GRB
by proton inelastic collisions with thermal nucleons \cite{bm00}. 
Proton acceleration definitely occurs in some sources, as evidenced by 
the detected cosmic rays.  Thus, the efficiency of neutrino generation 
in GRB \cite{wb97} and AGN \cite{muecke03}, two of the most widely 
suspected bottom-up astrophysical sources of cosmic rays, would give 
direct diagnostics for several of the key parameters relevant for 
the CR astrophysical acceleration hypothesis, as well as providing
crucial information on the physical conditions in these sources. Among 
such parameters are the baryon load of the (electromagnetically detected) 
jets, the injection rate and efficiency of proton acceleration in such 
jets, and the losses incurred (e.g. the density of target photons or 
nucleons, which translates into constraints on the magnetic field strenght, 
typical shock region dimension and jet bulk Lorentz factor).

The neutrinos, unlike the cosmic rays, point back at their source 
of origin; and unlike the photons, they will not be absorbed or 
obscured by intervening material. Aside from the  difficulty of 
detecting them, they constitute an ideal tomography probe of the
most dense and energetic regions of high energy sources. In the
TeV energy range, the expected angular resolution of cubic kilometer
ice or water Cherenkov detectors is $\theta\sim 0.5-1$ degree \cite{icecube},
which is well below the confusion limit. For burst-like sources, such 
as the 10-100 second duration GRB or the hour-day gamma-ray flares
in blazars, one expects both the angular and temporal signature to 
help drive the background down.

\begin{figure}
\includegraphics[width=16cm]{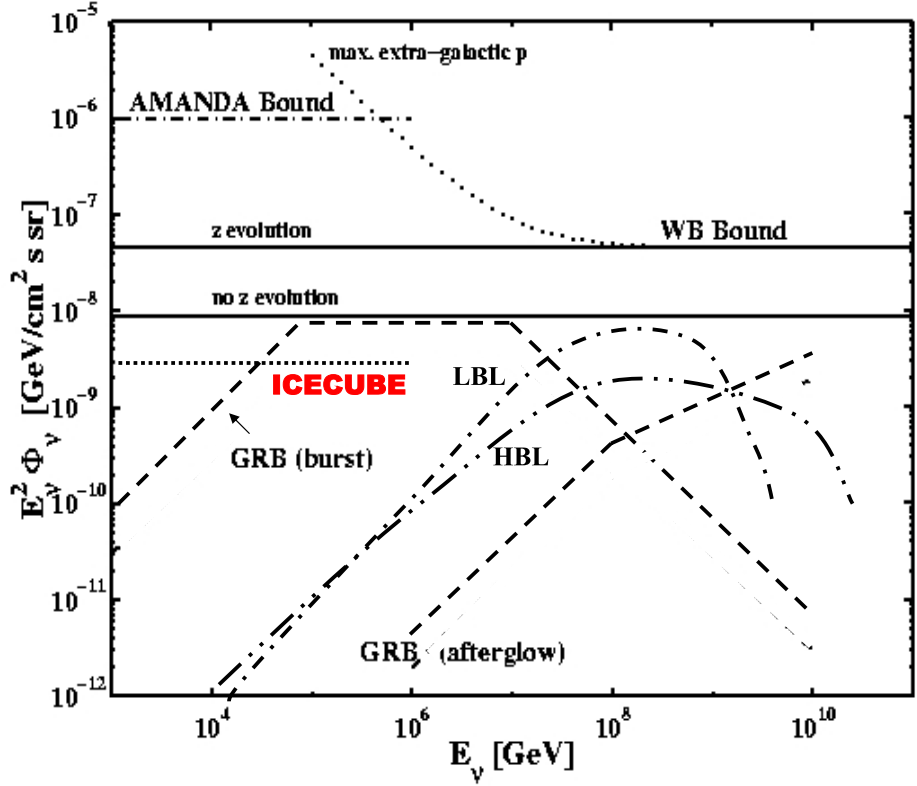}
\caption{The diffuse muon neutrino flux from proton acceleration
and $p,\gamma$ interactions in
a) GRB (burst) internal shocks and GRB afterglows, following 
\cite{rmw03b}, using current evolution and beaming constraints, 
without inclusion of a buried jet population; and
b) BL Lac galaxies including low peak blazars (LBL) and high peak
blazars (HBL), using the proton syncrotron blazar model \cite{muecke03}.
Also shown is the WB limit \cite{wb01}, and the atmospheric neutrino
background.
\label{fig:difnuspec}}
\end{figure}

One of the major questions in both AGN and GRB models is the
composition of the jets: are they purely MHD jets, dominated by
magnetic stresses and with an inertia provided mainly by $e^\pm$, or 
are they $e,p$ jets, where the inertia is mainly provided by baryons 
entrained in the jet? The astrophysical evidence is mixed. For AGN,
on the one hand the jet ram pressure inferred from the dynamics of 
the advance of the jet head into the intergalactic medium suggests
they are baryon loaded, while on the other hand, radio measurements
of the Faraday depolarization of the jet radio emission has suggested
in some cases that the jets have comparable numbers of electrons
and positrons.  Some degree of baryon entrainment is unavoidable, even 
if a jet is initially purely MHD, from exchange instabilities occurring 
at the interface between the jet and the galactic surroundings, so
that the question is really what is the degree of baryon loading.
For GRB, much recent excitement has been generated by the claim of
the detection of a high degree of gamma-ray linear polarization in 
the MeV range \cite{coburn03}. While debated, this observation could 
be suggestive of an MHD jet. On the other, baryonic jet models have 
proved much more useful in interpreting photon observations over a wide 
range of frequencies.

The dichotomy between leptonic (MHD) and hadronic jets has parallel
implications for the electromagnetic radiation of these objects, 
in particular for the TeV gamma-ray emission from very nearby blazars 
such as Mrk 421 and Mrk 501, which are at small enough redshifts to 
avoid excessive photopair attenuation by the diffuse infrared 
background. The usual leptonic jet interpretation of this radiation in 
terms of inverse Compton up-scattering of the synchrotron X-ray photons 
\cite{guetta04}
is certainly viable, and is viewed as a conservative extension of the 
sub-MeV phenomenology of well-studied high energy astrophysical sources.
The alternative hadronic jet interpretation relies on protons accelerated 
to energies $\gtrsim 10^5$ GeV leading to electromagnetic cascades and 
$e^\pm,\mu^+$ synchrotron and IC. This is also viable, although it 
requires denser photon target densities, and much higher magnetic fields 
than leptonic models, hence higher total energies \cite{aharonian00}. 
On the other hand, it could be viewed as conservative too,
in the sense that it is hard to see how protons could avoid being
accelerated, if present, and they would suffer much weaker losses during
the acceleration process.  At the simplest level, the test for both GRB 
and AGN is the prediction from purely MHD jets that they would produce no 
photomeson neutrinos, while the baryon-loaded $e,p$ (neutral) jets 
would produce $\sim 10-100$ TeV neutrino emission, as a consequence of 
the $p,\gamma$ interactions in the jets. These neutrinos are the least 
model dependent prediction, which arise typically, from 0.1-1 PeV 
protons interacting with X-rays in the jet comoving frame, whose bulk 
Lorentz factor are of order 10(100) for AGN (GRB). More model-dependent
predictions in GRB include EeV neutrinos from interaction of GZK energy
protons interacting with optical/UV photons arising in the reverse external
shock \cite{wb00}, and TeV neutrinos from proton interactios with thermal 
X-rays in pre-GRB buried jets \cite{mw01} making their way out through 
their massive stellar stellar progenitor's envelope. 

Some of the early AGN neutrino flux predictions have, in fact, already 
been indirectly constrained by the fact that they exceeded the so-called 
Waxman-Bahcall \cite{wb00} limit provided by the connection between utra-high 
energy cosmic rays and neutrinos. The predictions for less extreme AGN 
parameters, as well as the corresponding predictions for GRB, lie below this 
limit. The current sensitivity limit of AMANDA is just beginning to reach the 
level where it is comparable to the observed gamma-ray flux from a
specific AGN. Neutrino flux predictions involve assumptions about the 
proton to lepton ratio in the jets, as well as about the relative 
efficiency of injection of these particles into the acceleration process.
The latter is usually assumed to be some form of Fermi acceleration,
although in MHD models magnetic dissipation could lead to acceleration 
by the EMF of the transient electric fields. Explanations for the recent 
claim of a high gamma-ray polarization in a GRB have been  attempted
both in the context of MHD and baryonic jets, and a clear non-detection 
of 100 TeV neutrinos in GRB would lend some support to the former.

Observations of 10-100 TeV neutrinos in AGNs and GRBs, associated with 
GeV-TeV gamma-ray flares (e.g. observed with Whipple, Veritas, MILAGRO, 
HESS,  GLAST, etc. \cite{weekes04}) would provide convincing evidence for 
both a significant baryon content in the jets, and for efficient injection 
and acceleration of protons. 
This could rule out predominantly MHD jets, which in the case of GRB 
would put severe constraints on the magnetar scenario, where the 
central engine is assumed to be a strongly magnetized neutron star. It
could also constrain, both in GRB and AGN, the Blandford-Znajek mechanism 
for powering the jet by magnetic fields which couple to a fast-rotating
central black hole. A detection or non-detection would also constrain 
the location of the shocks, the photon energy density, the mechanism 
of production of the photons, and the efficiency for turbulent magnetic 
field amplification in the shocks.  The observation of EeV neutrinos,
implying GZK protons from GRB or AGN would require measuring extremely low
fluxes, possible only with experiments on the scale of EUSO/OWL. However 
the event rates of TeV neutrinos from buried pre-GRB jets is higher than
that of 100 TeV neutrinos coincident with the MeV gamma-rays, and 
would be observable with ICECUBE or ANTARES from individual bursts
a few times per decade. Micro-quasars, which are believed to be stellar-mass 
black hole accreting sources producing sem-relativistic jets, are less 
luminous but much closer in distance than GRB, and may also be detectable
individually in the TeV range. 

Such astrophysical studies are necessary to provide a base-line or
boundary, beyond which new physics may be considered compelling. 
Such measurements will allow to make novel tests of possible non-standard 
neutrino properties. For instance, neutrino decay would change the flavor 
ratios from the expected $\nu_e$:$\nu_\mu$:$\nu_\tau$=1:1:1; increases 
in the $\nu,N$ interaction at energies $\gtrsim 10^{18}$ eV due to black 
hole formation due to extra dimensions, or tachyonic effects, would give 
substantially greater fluxes than the modest ones predicted by standard 
model astrophysics, etc.


\newpage
\begin{center}
{\Large\sc Dark matter searches using neutrinos}\\
\end{center}

\newcommand{\postscript}[2]{\setlength{\epsfxsize}{#2\hsize}
   \centerline{\epsfbox{#1}}}
\newcommand{\mweak}{M_{\text{Weak}}}
\newcommand{\mplanck}{M_{\text{Pl}}}
\renewcommand{\mp}{M_p}
\newcommand{\mstar}{M_{*}}
\newcommand{\md}{M_D}
\newcommand{\ifb}{\text{fb}^{-1}}
\newcommand{\ev}{\text{eV}}
\newcommand{\mev}{\text{MeV}}
\newcommand{\gev}{\text{GeV}}
\newcommand{\tev}{\text{TeV}}
\newcommand{\pb}{\text{pb}}
\newcommand{\mb}{\text{mb}}
\newcommand{\cm}{\text{cm}}
\newcommand{\km}{\text{km}}
\newcommand{\g}{\text{g}}
\newcommand{\s}{\text{s}}
\newcommand{\yr}{\text{yr}}
\newcommand{\sr}{\text{sr}}
\newcommand{\etal}{{\em et al.}}
\newcommand{\eg}{{\em e.g.}}
\newcommand{\ie}{{\em i.e.}}
\newcommand{\ibid}{{\em ibid.}}
\newcommand{\eqref}[1]{Eq.~(\ref{#1})}
\newcommand{\eqsref}[2]{Eqs.~(\ref{#1}) and (\ref{#2})}
\newcommand{\figref}[1]{Fig.~\ref{fig:#1}}
\newcommand{\sla}[1]{\not{\! #1}}

\newcommand{\sign}{\:\!\text{sgn}\:\!}
\newcommand{\mgaugino}{M_{1/2}}
\newcommand{\mt}{m_t}
\newcommand{\mgut}{M_{\text{GUT}}}
\newcommand{\tb}{\tan\beta}
\newcommand{\ethr}{E_{\text{thr}}}
\newcommand{\eopt}{E_{\text{opt}}}
\newcommand{\mchi}{m_{\chi}}
\newcommand{\Omegachi}{\Omega_{\chi}}
\newcommand{\OmegaDM}{\Omega_{\text{DM}}}
\newcommand{\be}{\begin{equation}}
\newcommand{\ee}{\end{equation}}

\newcommand{\met}{\rlap{\,/}E_T}
\newcommand{\gsim}{ \mathop{}_{\textstyle \sim}^{\textstyle >} }
\newcommand{\lsim}{ \mathop{}_{\textstyle \sim}^{\textstyle <} }

\newcommand{\mB}{m_{B^1}}
\newcommand{\mq}{m_{q^1}}
\newcommand{\mf}{m_{f^1}}
\newcommand{\mKK}{m_{KK}}

\newcommand{\bold}[1]{{\text{\normalsize\bm{$#1$}}}}
\newcommand{\rem}[1]{{\bf #1}}


In recent years there has been tremendous progress in our
understanding of the universe on the largest scales.  For the first
time, cosmological measurements have provided a complete census of the
universe.  In units of the critical density, the energy densities of
baryons, non-baryonic dark matter, and dark energy
are~\cite{Spergel:2003cb,Tegmark:2003ud}
\begin{eqnarray}
\Omega_B &=& 0.044 \pm 0.004  \\
\Omega_{\text{DM}} &=& 0.23 \pm 0.04 \label{Omegadm} \\
\Omega_{\Lambda} &=& 0.73 \pm 0.04 \ .
\end{eqnarray}
At the same time, the microscopic identities of dark matter and dark
energy are at present unknown and are among the most important open
questions in science today.  New particles are required, and a
fundamental understanding of the dark universe therefore draws on
complementary approaches from both cosmology/astrophysics and
particle/nuclear physics.

Neutrinos play a unique and promising role in resolving these
mysteries.  This is especially true in the case of dark matter, where
the importance of neutrinos may be understood from simple and general
considerations.  The stability of individual dark matter particles is
typically guaranteed by a conserved parity.  These conservation laws,
however, allow pairs of dark matter particles to annihilate into
ordinary particles, providing a signal for dark matter detection.
Such signals are, of course, greatly enhanced when the dark matter
particle density and annihilation rate are large, as they are expected
to be at the center of astrophysical bodies.  Unfortunately, when dark
matter particles annihilate in these regions, most of their
annihilation products are immediately absorbed.  Neutrinos, however,
are not.  High energy neutrinos from the cores of the
Sun~\cite{Gaisser:1986ha,Press:1985ug,Silk:1985ax,%
Srednicki:1987vj,Hagelin:1986gv,Ng:1987qt,Ellis:1988sh} and
Earth~\cite{Freese:1986qw,Krauss:1986aa,Gaisser:1986ha,Gould:1989eq}
are therefore promising signals for dark matter detection.

The neutrino flux from dark matter annihilation is determined by a
number of factors.  First and foremost, it depends on the dark matter
number density at the source, which is governed by the competing
processes of capture and annihilation.  A dark matter particle $\chi$
is captured when an interaction $\chi N \to \chi N$ reduces its
velocity to below the escape velocity.  Once this happens, subsequent
interactions typically allow the dark matter to settle to the core. At
the same time, dark matter particles annihilate in the core through
the processes $\chi \chi \to f \bar{f}, WW, ZZ$, reducing the number
of $\chi$ particles.

If $C$ is the capture rate and $A$ is the total annihilation cross
section times relative velocity per volume, the number $N$ of dark
matter particles at the source satisfies $\dot{N} = C - A N^2$.  The
present annihilation rate is therefore
\begin{equation}
\Gamma_A = \frac{1}{2} A N^2 = \frac{1}{2} C \tanh^2(\sqrt{CA}\, t)\ ,
\end{equation}
where $t$ is the collection time. For signals from the Sun or Earth,
$t \approx 4.5$ Gyr, the age of the solar system.  For large enough
$t$, the annihilation rate approaches its maximal value $\Gamma_A =
\frac{1}{2} C$ and is a function of the capture rate alone. More
generally, however, the neutrino flux depends on both $C$ and $A$, in
contrast to direct detection rates, which depend only on scattering
cross sections, and other indirect detection rates, which depend only
on the annihilation cross sections.

Neutrinos produced in the annihilation processes $\chi \chi \to f
\bar{f}, WW, ZZ$ propagate to the Earth's surface, where they may be
detected through their charged-current interactions.  The most
promising signal is from upward-going muon neutrinos that convert to
muons in the surrounding rock, water, or ice, producing through-going
muons in detectors.  The detection rate for such neutrinos is greatly
enhanced for high energy neutrinos, as both the charged-current cross
section and the muon range are proportional to $E_{\nu}$.

The resulting muon fluxes are sensitive to many effects and are
subject to astrophysical uncertainties.  For standard halo dark matter
populations, the signal is fairly well-determined, as it depends
primarily on the local dark matter density, and so is insensitive to
details of halo models.  Additional dark matter populations may,
however, significantly enhance predictions for muon fluxes.  (See, for
example, Refs.~\cite{Bergstrom:1999tk,Gould:1999je}.)  Muon signals
are also sensitive to the details of capture rates~\cite{Gould:1987ju}
and effects in propagating the neutrinos from the core to the
surface~\cite{Gaisser:1986ha,Ritz:1988mh,Jungman:1995jr,Crotty:2002mv}.

The neutrino signal for dark matter detection has been analyzed for a
variety of dark matter possibilities.  Among the most compelling
candidates are WIMPs, weakly-interacting massive particles. Such
particles have masses given by the weak scale $M_W \sim {\cal
O}(100)~\gev$, and interact with ordinary matter with cross sections
$\sigma \sim M_W^{-4}$.  WIMPs are motivated not only by the dark
matter problem, but also independently by attempts to understand the
electroweak scale and electroweak symmetry breaking.  Even more
tantalizing, the thermal relic density of dark matter particles
emerging from the hot Big Bang may be calculated given their mass and
interaction cross sections.  For the typical WIMP parameters given
above, the relic density is naturally in a range consistent with the
observed value of \eqref{Omegadm}.

The prototypical WIMPs, and by far the most studied, are neutralinos
in supersymmetry.  Neutralinos are Majorana fermions that are, in
general, mixtures of the fermionic superpartners of the U(1)
hypercharge gauge boson, the neutral SU(2) gauge boson, and the Higgs
scalars.  Their Majorana-ness has an immediate implication for the
neutrino signal.  For a pair of Majorana fermions, the initial state
has spin 0, and so the most promising possible signal, $\chi \chi \to
\nu \bar{\nu}$, is effectively absent, since it is helicity- or
$P$-wave-suppressed.  As a result, for neutralino dark matter, the
highest neutrino energies are typically $E_{\nu} \sim
\frac{1}{2}\mchi$ from $\chi \chi \to WW, ZZ$ to $\frac{1}{3}\mchi$
from $\chi \chi \to \tau \bar{\tau}$.

\begin{figure}
\includegraphics[width=6truein]{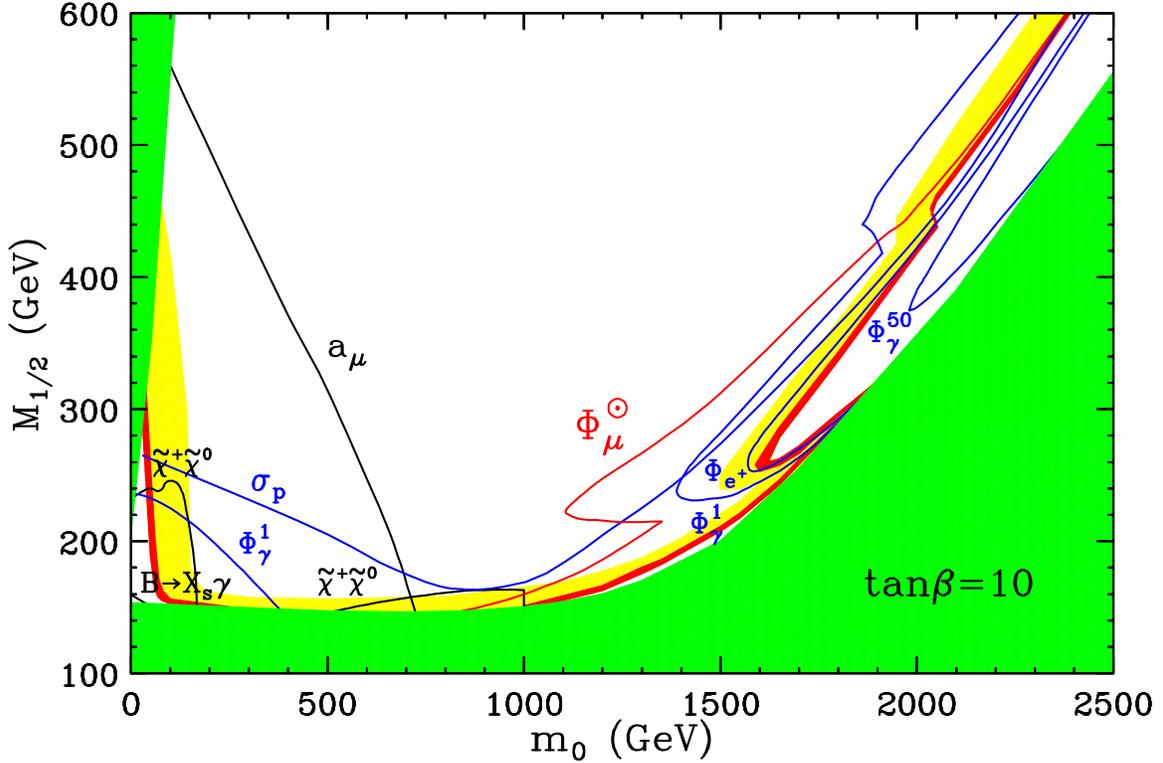}
\caption{Estimated reaches of neutrino telescope searches for
neutralino dark matter ($\Phi_{\mu}^{\odot}$, red), other dark matter
searches (blue), and various high-energy collider and low-energy
precision searches (black) in minimal supergravity parameter space.
The remaining minimal supergravity parameters are fixed to $\tb=10$,
$A_0 = 0$, and $\mu > 0$. The green shaded regions are excluded by
chargino mass bounds and the requirement that the dark matter particle
be neutral.  The regions probed extend the labeled contours toward the
forbidden, green regions.  In the red and yellow shaded regions, the
neutralino thermal relic density satisfies post-WMAP ($0.094 <
\OmegaDM h^2 < 0.129$) and pre-WMAP ($0.1 < \OmegaDM h^2 < 0.3$)
bounds, respectively.  Updated from Ref.~\protect\cite{Feng:2000zu}.
\label{fig:reach10}}
\end{figure}

Given a particular supersymmetric model framework, the reach of
neutrino telescopes for discovering neutralino dark matter may be
determined.  The results for minimal supergravity are given in
\figref{reach10}.  Minimal supergravity is a simple framework that
encapsulates many appealing features of supersymmetry.  It assumes
that the supersymmetric scalars and gauginos have unified masses $m_0$
and $\mgaugino$ at the scale of force unification.  The reach of
neutrino telescopes in the $(m_0, \mgaugino)$ plane is shown in
\figref{reach10}.  The region probed by neutrinos from the Sun is
indicated by the $\Phi_{\mu}^{\odot}$ contour, where a sensitivity to
muon fluxes above $100~\km^{-2}~\yr^{-1}$ is assumed.  Such
sensitivities may be reached in the near future by experiments such as
AMANDA and ANTARES.  These and other neutrino telescopes, along with
their more salient characteristics and flux limits (where available),
are listed in Table~\ref{table:NT}.

\begin{table}[tb]
\caption{Current and planned neutrino experiments.  We list also each
experiment's start date, physical dimensions (or approximate effective
area), muon threshold energy $E_{\mu}^{\rm thr}$ in GeV, and 90\% CL
flux limits for the Earth $\Phi_{\mu}^{\oplus}$ and Sun
$\Phi_{\mu}^{\odot}$ in $\km^{-2}~\yr^{-1}$ for half-cone angle
$\theta \approx 15^{\circ}$ when available.  {}From
Ref.~\protect\cite{Feng:2000zu}.
\label{table:NT} }
\begin{center}
\begin{tabular}{lllrrrr}
 Experiment \rule[-2.8mm]{0mm}{7mm}
 & \hspace*{3mm} Type
  & Date
   & Dimensions\hspace*{2mm}
    & $E_{\mu}^{\rm thr}$
     & $\Phi_{\mu}^{\oplus}$\hspace*{2mm}
      & $\Phi_{\mu}^{\odot}$\hspace*{2mm}  \\  \hline
 Baksan
 & Ground
  & 1978
   & $17 \times 17 \times 11 ~\text{m}^3$
    & 1  
     & $6.6 \times 10^3$ 
      & $7.6 \times 10^3$ \\  
 Kamiokande
 & Ground
  & 1983 
   & $\sim 150 ~\text{m}^2$
    & 3
     & $10 \times 10^3$ 
      & $17 \times 10^3$ \\
 MACRO
 & Ground
  & 1989
   & $12 \times 77 \times 9 ~\text{m}^3$
    & 2
     & $3.2 \times 10^3$ 
      & $6.5 \times 10^3$ \\  
 Super-Kamiokande
 & Ground
  & 1996
   & $\sim 1200~\text{m}^2$
    & 1.6  
      & $1.9 \times 10^3$ 
       & $5.0 \times 10^3$  \\  
 Baikal NT-96
 & Water
  & 1996
   & $\sim 1000~\text{m}^2$
    & 10  
      & $15 \times 10^3$ 
       &              \\  
 AMANDA B-10
 & Under-ice
  & 1997 
   & $\sim 1000~\text{m}^{2\, \dag}$\hspace*{-1.73mm}
    & $\sim 25$
     & $44 \times 10^{3\, \dag}$\hspace*{-1.73mm}
      &     \\
 Baikal NT-200
 & Water
  & 1998
   & $\sim 2000~\text{m}^2$
    & $\sim 10$  
     &  \\  
 AMANDA II
 & Ice
  & 2000
   & $\sim 3 \times 10^4~\text{m}^2$
    & $\sim 50$
     & 
      &     \\
 NESTOR$^{\S}$
 & Water
  & 
   & $\sim 10^4~\text{m}^{2\, \ddag}$\hspace*{-1.73mm}
    & few
     &
      &     \\
 ANTARES
 & Water
  & 
   &  $\sim 2 \times 10^4~\text{m}^{2\, \ddag}$\hspace*{-1.73mm}
    & $\sim 5$--10
     &
      &     \\    
 IceCube
 & Ice
  & 
   & $\sim 10^6~\text{m}^2$
    & 
     & 
      &    
\end{tabular}
\end{center}

$^\dag$ Hard spectrum, $\mchi = 100~\gev$. \
$^{\S}$ One tower. \
$^\ddag$ $E_{\mu} \sim 100~\gev$.
\end{table}

Figure~\ref{fig:reach10} illustrates many key points.  First, the
reach of neutrino telescopes is significant.  In \figref{reach10} the
red shaded region indicates the region of parameter space in which the
neutralino thermal relic density is in the range favored by WMAP.
Although current bounds do not restrict the parameter space
substantially, the neutrino signal is observable in the near future
for all of the red region with $m_0 \agt \tev$.  Second, the neutrino
signal, along with other indirect dark matter signals, is
complementary to other signals of supersymmetry.  The reaches of many
other supersymmetry searches are also shown in \figref{reach10}.  We
find that traditional particle physics signals, such as the trilepton
signal from $\chi^{\pm} \chi^0$ production at the Tevatron or
deviations in $B \to X_s \gamma$ and the muon's anomalous magnetic
moment $a_{\mu}$, are effective for $m_0 \alt \tev$, but are
ineffective for large $m_0$.  Neutrino telescopes therefore provide a
complementary method for searches for new physics.

Muon energy thresholds, listed in Table~\ref{table:NT}, have not been
included in \figref{reach10}.  Since the muon detection rate is
dominated by high energy muons as noted above, the threshold energy is
typically not important, especially in the regions where a detectable
signal is expected.  This is not the case for all detectors, however.
For example, since muons lose $0.26~\gev$ per meter in water and ice,
neutrino telescopes requiring track lengths of $\sim 100$ m will have
thresholds of order $\sim 25~\gev$. The dependence on threshold energy
has been studied in Refs.~\cite{Bergstrom:1997tp,Bergstrom:1998xh},
where it was found that for threshold energies of $E_{\mu}^{\rm thr}
\sim \frac{1}{4}\mchi$ to $\frac{1}{6}\mchi$, the loss of signal is
substantial.  Low threshold energies in neutrino telescopes are
clearly very important for dark matter detection.  

Several other studies of neutrino telescope dark matter searches have
been conducted in this and more general supersymmetric
frameworks~\cite{Hooper:2003ka}, and also in completely different
scenarios.  A particularly interesting example is the case of theories
with extra spatial dimensions.  In so-called universal extra dimension
scenarios~\cite{Appelquist:2000nn}, in which all fields of the
standard model propagate in all extra dimensions, the lightest
Kaluza-Klein particle may be the dark matter.  The requirement that
the low energy degrees of freedom include only those of the standard
model typically imposes a Kaluza-Klein parity, which makes the
lightest Kaluza-Klein particle stable, and for $\tev^{-1}$ size extra
dimensions, the lightest Kaluza-Klein particle is a WIMP with
naturally the correct thermal relic density~\cite{Servant:2002aq}.

If $B^1$, the lightest Kaluza-Klein partner of the hypercharge gauge
boson, is the dark matter, the dark matter is a massive spin 1
particle.  Angular momentum constraints, which eliminated the most
promising models in the neutralino case, then do not imply helicity
suppression for annihilation to fermion pairs.  The annihilation $\chi
\chi \to \nu \bar{\nu}$ is therefore unsuppressed, and provides, in
fact, the dominant source of neutrinos, with $E_{\nu} = \mchi$.
Studies of the neutrino signal for Kaluza-Klein dark
matter~\cite{Cheng:2002ej,Hooper:2002gs,Bertone:2002ms} have found
reaches up to $m_{B^1} = 600~\gev$ for AMANDA and ANTARES, and up to
$m_{B^1} = 1.4~\tev$ for IceCube, competitive with other methods for
discovering dark matter in these scenarios.


\newpage
\begin{center}
{\Large\sc Neutrinos as a probe of supernovae}\\
\end{center}


The deaths of massive stars through spectacular stellar explosions
known as core collapse supernovae are the single most important source
of elements in the Universe. They are the dominant source of elements
between oxygen and iron and are believed to be responsible for
producing half the elements heavier than iron. Many of these elements
were necessary for the evolution of life. Moreover, gamma-ray and
x-ray observations of gamma-ray bursts have now illuminated an
association between these bursts and core collapse supernovae, and in
the last five years a number of hyper-energetic core collapse
supernovae have been observed, reaffirming their status as the most
energetic events in the Universe. And we now stand at a
threshold. Galactic core collapse supernovae are among the sources
expected to generate gravitational waves that can be detected by
observatories around the world. A detection would mark the birth of an
entirely new subfield of astronomy and would be the first direct
evidence of the physical and dynamic nature of spacetime.

In addition to their place in the cosmic hierarchy, the extremes of
density, temperature, and composition encountered in core collapse
supernovae provide an opportunity to explore fundamental nuclear and
particle physics that would otherwise be inaccessible in terrestrial
experiments. Supernovae serve as cosmic laboratories, and supernova
models are the bridge between observations (bringing us information
about these explosions) and the fundamental physics we seek. In the
event of a Galactic core collapse supernova, the joint detection of
the neutrinos and gravitational waves would provide a wealth of
information that will both help us develop and validate supernova
models and, given sufficiently sophisticated models, allow us to
extract significant information about the high-density, neutron-rich
environment in the collapsing stellar core and proto-neutron star.

The connections between supernova observations, supernova models, and
fundamental physics is brought full circle when we consider the role
of terrestrial nuclear physics experiments.  For example, measurements
of neutrino-nucleus cross sections validate the nuclear structure
models that underpin the calculations of stellar core neutrino-nucleus
weak interactions. These define the dynamics of stellar core collapse
and set the stage for the supernova dynamics that occurs after stellar
core bounce \cite{hix03}. Terrestrial experiments will enable more
sophisticated supernova models, which in turn will allow, when
detailed neutrino observations become available, a more accurate
discrimination of nuclear models for the high-density, neutron-rich
stellar core regions.

Core collapse supernovae result from the gravitational collapse of the
iron core of a massive star (more than ten times the mass of the
Sun). The inner iron core undergoes a transition to nuclear matter and
rebounds at supernuclear densities to launch a shock wave into the
outer iron core. This shock wave dissociates the outer iron core
nuclei to form a hot mantle of neutrons and protons outside of the
unshocked inner, nuclear matter core. The shock wave must ultimately
propagate out of the iron core and through the successive stellar
layers of increasingly lighter elements to produce the supernova.

Much work remains to be done to elucidate the core collapse supernova
mechanism. These are neutrino-driven events involving the turbulent
fluid flow in the exploding stellar core, stellar core rotation and
magnetic fields, and strong (general relativistic) as opposed to weak
(Newtonian) gravitational fields. Meeting the scientific challenge
will ultimately require three-dimensional general relativistic
radiation magnetohydrodynamics simulations with three-dimensional,
multi-neutrino-energy, multi-neutrino-angle neutrino transport. This
state of the art macrophysics must be matched by state of the art
microphysics that will describe the stellar core sub- and
super-nuclear density nuclear physics and the neutrino-stellar core
weak interaction physics. The cooperation of nuclear physicists,
particle physicists, and astrophysicists has already proven to be a
very effective way to address this Grand Challenge problem (e.g., see
the DOE SciDAC-funded TeraScale Supernova Initiative:
www.tsi-scidac.org), and a sustained large-scale, multi-physics effort
will be needed to systematically arrive at its solution. Current
supercomputers are enabling the first detailed two-dimensional
simulations. A ten-fold increase in capability would provide, for the
first time, the opportunity to simulate supernovae realistically, in
three dimensions. This increase in computing power is expected within
the next five years. Given the challenge of modeling such nonlinear
multi-physics systems, and the opportunity afforded by Tera- and
Peta-Scale computing platforms, input from experiment and/or
observation is both essential and timely.

The neutrino and gravitational wave emissions from core collapse
supernovae provide direct information about the dynamics at the center
of the exploding star and, therefore, about the explosion mechanism
itself. Instabilities in the proto-neutron star, such as convection or
doubly-diffusive instabilities (e.g., lepto-entropy fingers, neutron
fingers), and rotation will have a direct impact on the emergent
neutrino fluxes and, therefore, the explosion mechanism and supernova
byproducts. Thus, a detailed three-flavor neutrino pulse detection by
SK, SNO, ICARUS, KamLAND, and LVD, among others, would provide a
template against which models of the turbulent and rotating stellar
core dynamics could be developed and validated.

The neutrino emission from a core collapse supernova occurs in three
major stages: (1) The emission of electron neutrinos during stellar
core collapse and an electron neutrino burst $\sim 10^{53}$ erg/s
only milliseconds after stellar core bounce, as the supernova shock
wave passes through the electron neutrinosphere, at which point the
trapped electron neutrinos behind the shock generated by rapid
electron capture on the shock-liberated protons are free to
escape. (2) A longer-term Kelvin-Helmholtz cooling phase of the
proto-neutron star, which occurs over a period $\sim$10 s during which
time the supernova is launched and the proto-neutron star cools to
form a neutron star or a black hole. This phase is characterized by
the emission of neutrinos of all three flavors, and their
antineutrinos, from the hot, proto-neutron star mantle and the
liberation in the form of these neutrinos of the $\sim 10^{53}$ erg of
gravitational binding energy of a neutron star. The neutrinos are
emitted at the staggering rate $\sim 10^{52-53}$ erg s$^{-1}$ with RMS
energies between 10 and 25 MeV. (3) A long-term ($\sim$10-100 s)
neutron star cooling phase.

The electron neutrino burst is a probe of the physics of stellar core
collapse, bounce, and initial shock formation and propagation
\cite{tbp03}. It would be observable in the event of a Galactic
supernova. Electron capture on nuclei during stellar core collapse
determines the size of the inner iron core and thus the initial
location and energy of the supernova shock wave at core bounce, which
then sets the stage for everything that follows. In turn, electron
capture on the nuclei in the stellar core during collapse depends on
their detailed nuclear structure. An electron-neutrino burst detection
has never been achieved. Such a detection would shed light on both
stellar core collapse and nuclear structure physics.

Regarding the second neutrino emission phase, current supernova models
indicate that a combination of fluid instabilities below the supernova
shock wave and rotation of the stellar core will affect the
luminosities and mean energies of the neutrinos emanating from the hot
mantle during this phase \cite{fh00,buras03}. To complicate matters
further, in the case of a rotating core the observed neutrino
luminosities along the rotation axis and along the equator could
differ by as much as a factor of three \cite{jm89}.

Current models also indicate that the stellar core collapse and bounce
dynamics, which is tied to the rotation of the core, and the different
fluid instabilities that may develop below the shock will have
distinct gravitational wave signatures that, in the event of a
Galactic supernova, may be detectable by both LIGO I and LIGO II
\cite{mueller04}. The gross stellar core collapse and bounce and the
different fluid instabilities will yield gravitational wave signatures
at different frequencies in the 10-3000 Hz range, and with different
amplitudes. Thus, core collapse supernovae provide an opportunity for
a joint detection of both neutrinos and gravitational waves, with one
leveraging the other; each would provide a complementary diagnostic
for supernova models.

But all of the models mentioned above are far from complete. In
reality, the impact of a detailed neutrino (and gravitational wave)
detection would be felt at an even more fundamental level. Currently,
there are no core collapse supernova models that both implement
realistic neutrino transport and explode, whether the models are
one-dimensional (spherically symmetric)
\cite{rj00,liebendoerfer01,tbp03} or two-dimensional (axisymmetric)
\cite{mezzacappa98,buras03} (there are as yet no three-dimensional
models with sufficiently realistic neutrino transport). Moreover,
there are no sufficiently complex models that include all of the known
potentially relevant physics and none at all that include magnetic
fields and are sufficiently realistic in other respects that can
reliably explore the possible role magnetic fields may play in
supernova dynamics \cite{akiyama03}. Thus, a detailed neutrino light
curve would illuminate much about the nature of stellar core collapse
and bounce, the postbounce evolution, and the physics responsible for
the generation of the supernova.

Detailed neutrino signatures will supply volumes of information not
only about the macroscopic physics of the stellar core during
explosion but about its microscopic physics. In particular, emergent
neutrino fluxes will depend on the energetics of shock formation and
the proto-neutron star evolution, both of which depend on the
high-density equation of state at and after bounce. For example, the
development of fluid instabilities in the proto-neutron star depend on
this equation of state \cite{brm04}. In turn, these affect the
neutrino fluxes, with vigorous instabilities potentially associated
with significant boosts in luminosity. At late times, during phase
three of the neutrino signature from core collapse supernovae,
compositional changes (e.g., the existence of quarks) can have
catastrophic consequences, leading to metastability of the newly
formed neutron star and the dynamic collapse to a black hole, with a
corresponding sharp cutoff in the neutrino luminosities. Such events
are potentially detectable for a Galactic supernova, especially if
detector sizes are significantly increased and/or new detector
technologies are implemented~\cite{bh00,bh01}.

It is exciting to think about the scientific revolution that would
occur in the event of a Galactic supernova, but at the same time one's
excitement is always tempered by the fact that, on average, a Galactic
core collapse supernova is expected only twice per century (although
the birth rate of pulsars, which are born in core collapse supernovae,
has recently been estimated to be at the significantly higher rate of
four per century \cite{c04}). On the other hand, if such supernovae
were visible in neutrinos out to the Virgo cluster, the core
collapse supernova rate would rise dramatically to a few per
year. Thus, both larger-scale experiments and new neutrino detection
technologies, which would make this possible, should seriously be
explored. The scientific payoff would be substantial.

Finally, it is now an experimental fact that neutrinos have mass. The
implications are far-reaching and include implications for core
collapse supernova dynamics and phenomenology. First, neutrino mixing
deep within the stellar core could have a dramatic impact on the
emergent neutrino fluxes from the proto-neutron star and thereby
affect the explosion mechanism and observables \cite{sf02}. Second,
neutrino mixing also leads to characteristic signatures in supernova
neutrino detection.  Consequently,
knowledge and deconvolution of this mixing would be required in order
to use the detected fluxes to validate supernova models and extract
information relating to the high-density nuclear physics. Thus,
developments in supernova theory and our ultimate understanding of
these great cosmic events is not independent of exciting developments
that have and will take place in other areas of astrophysics and
particle physics.


\newpage
\begin{center}
{\Large\sc Supernova neutrinos as tests of particle physics}\\
\end{center}


Core collapse supernovae are widely touted as laboratories for 
fundamental neutrino, nuclear, and particle physics. In fact, there are 
aspects of this assertion which are true and aspects which are false. 
The truth is that much of the physics and phenomenology of supernova 
core collapse and explosion {\it is} sensitive to fundamental issues in 
the weak interaction, especially as regards neutrinos. However, it is 
frequently also true that we do not yet understand the core collapse 
supernova phenomenon at a level sufficient to turn this sensitivity to 
input physics into hard constraints on fundamental particle physics.

Aiding and abetting our ignorance of the details of core collapse 
supernovae is the sad state of supernova neutrino astronomy: our only 
record of the neutrino signal from collapse is the handful of events 
from SN 1987a. There is, on the other hand, every reason to think that 
progress will be made and that a better understanding of core collapse 
physics, perhaps coupled with a neutrino signature in modern detectors 
from a Galactic collapse event, will allow us to exploit the tremendous 
and tantalizing sensitivity of collapse/explosion physics to neutrino 
properties, flavor mixing, as well as new physics. It is even 
conceivable that new fundamental neutrino physics could be a key to 
understanding, for example, why (some?) collapse events lead to 
explosions and perhaps how heavy elements are formed.

Why are core collapse supernovae so sensitive to input weak interaction 
and neutrino physics and yet so difficult to understand? We can give a 
succinct answer to this: (1) nearly all of the energy available in this 
problem resides in seas of neutrinos in the core; (2) the energy of the 
supernova explosion is tiny, comprising only some $1\%$ of the energy 
in these neutrino seas. The essence of the supernova problem is to 
figure out how $\sim 1\%$ of the energy resident in the neutrino seas 
in the core is transported out and deposited behind or near a nascent 
shock. The transport problem depends, in turn, on uncertain aspects of 
the high density nuclear equation of state, weak interactions at high 
density, and perhaps multi-dimensional issues in radiation 
hydrodynamics coupled with neutrino flavor evolution physics. Magnetic 
fields and MHD may play a role as well. And in all of this great 
accuracy is necessary because, for example, the explosion energy is 
such a small fraction of the total energy.

With this discouraging assessment particle physicists may be heading 
for the exits. Not so fast. The possibilities are seductive, however, 
as this whole collapse/explosion process can be exquisitely sensitive 
to lepton number violation. The collapse/explosion environment affords 
a unique sensitivity to neutrino interactions and properties that are 
matched nowhere else in the universe save, possibly, for the Big Bang.

When the Chandrasekhar mass iron core collapses to a distended neutron 
star in $\sim 1\,{\rm s}$ some $10^{52}\,{\rm ergs}$ of gravitational 
binding energy are released promptly, roughly $99\%$ of this as 
neutrinos of all flavors. This complements the sea of mainly $\nu_e$ 
neutrinos produced by electron capture during the collapse or Infall 
Epoch. The collapse is halted at or above nuclear density and a shock 
wave is generated. This shock moves out to hundreds of kilometers from 
the core on a timescale of order a hundred milliseconds and stalls. 
Meanwhile, and subsequently, the neutron star core contracts 
quasi-statically, eventually releasing some $10^{53}\,{\rm ergs}$ as 
neutrinos. The energy in the neutrino seas at this point is $\sim 10\%$ 
of the rest mass of the neutron star!

All aspects of this process are sensitive to neutrino flavor evolution. 
If $\nu_e$'s are converted to another type of neutrino, either active 
or sterile, then electron capture will be unblocked, the electron 
fraction will be reduced (rendering the collapsing stellar core more 
neutron-rich), and the subsequent bounce shock will be weaker and the 
thermodynamic structure of the core will be altered, in broad brush 
increasing the entropy for active-active transformation and likely 
decreasing it for active-sterile conversion. The weaker shock energy 
will translate into a smaller radius of stall-out and, perhaps, a 
reduced re-heating rate from neutrino energy deposition. If we 
understood where the explosion came from we might be able to use these 
considerations to put rather stringent constraints on a number of 
lepton number violating interactions/processes in the Infall Epoch.

Clearly, a better understanding of the shock re-heating process would 
be beneficial.  We could someday get this from a detailed neutrino 
signal. Likewise, it is conceivable that nucleosynthesis, especially of 
the heavy r-process elements, could give us important clues about what 
is going on after core bounce. Many neutrino processes which involve 
lepton number violation can be important in this post-bounce regime.

Active-sterile neutrino flavor conversion has been invoked over a wide 
range of sterile neutrino masses and mixings to, among other things, 
explain pulsar kicks and enhanced shock re-heating, and to give a 
robust scenario for successful r-process nucleosynthesis in slow 
neutrino-driven winds. The neutrino process (neutrino-nucleus 
spallation) may also be sensitive to this process if active neutrino 
energies are increased.

Obviously, the active neutrino signal from the supernova can also be 
affected by both active-active and active-sterile neutrino flavor 
conversion. The role of the density jump associated with the shock can 
be important. In fact, it may afford insight into the neutrino mass and 
mixing spectrum, shedding light on, for example, $\theta_{13}$. 
Alternatively, if these neutrino mixing properties are known 
independently from laboratory experiments, then the neutrino signal 
could let us probe shock propagation during and after the re-heating 
event.

In all of these regimes, new neutrino interactions, such as those from 
extensions of the Standard Model can have important effects. Our model 
for collapse and explosion could be changed by new interactions, though 
perhaps not yet to a large enough extent for legitimate constraint. The 
history of the supernova problem makes this abundantly clear. For 
example, the discovery of neutral currents completely altered our view 
of stellar collapse.

Finally, suppose we restrict our attention to the known active 
neutrinos. The recent experimental revolution in neutrino physics has 
arguably given us all the neutrino mass-squared differences (though not 
the absolute masses) and all of the vacuum mixing properties save for 
$\theta_{13}$ and the CP-violating phase. The unfortunate truth is that 
we are not yet able to take this hard-won data and calculate the 
consequences for neutrino flavor evolution either inside or above the 
neutron star! This is because the neutrino flavor evolution problem in 
the supernova environment, unlike the sun, is fiercely non-linear and 
in a unique way. The potential which governs neutrino flavor 
transformation is dominated by neutrino-neutrino forward scattering. In 
turn, this process depends on the flavor states of the neutrinos.

If we consider the coherent evolution/propagation of neutrinos above 
the neutron star, then a given neutrino world line will intersect the 
world lines of other neutrinos and, possibly, lead to a forward 
scattering events. Essentially all such trajectories have flavor 
evolution histories which are coupled to one another through quantum 
entanglement engendered by forward scattering. This is a unique kind of 
transport problem.

Of course, if the emergent fluxes and energy spectra of the various 
flavors of neutrinos at the neutron star surface are identical, then 
swapping neutrino flavors will have no effect on any aspect of 
supernova physics. In fact, current simulations and neutrino transport 
calculations seem to suggest that this is the case at early times 
post-bounce. Whether or not this remains true over the some $20\,{\rm 
s}$ of neutron star contraction where neutrino fluxes are appreciable 
remains to be determined. Changes in the core, especially to the 
equation of state and to the net lepton number residing in the core, 
likely will lead to energy spectrum and flux differences between the 
neutrino flavors.

In any case, progress on understanding the core collapse supernova 
phenomenon may well depend on what we know about and our ability to 
model how the weakly interacting sector evolves. The daunting nature of 
the supernova problem should not discourage us from striving for better 
insight.


\newpage
\begin{center}
{\Large\sc Diffuse supernova neutrino background}\\
\end{center}

\def\la{\mathrel{\mathpalette\fun <}}
\def\ga{\mathrel{\mathpalette\fun >}}
\def\fun#1#2{\lower3.6pt\vbox{\baselineskip0pt\lineskip.9pt
  \ialign{$\mathsurround=0pt#1\hfil##\hfil$\crcr#2\crcr\sim\crcr}}}
\def\eg{{\it e.g.}}
\def\ie{{\it i.e.}}
\def\etal{{\it et al.~}}
\newcommand{\snr}{{\rm N_{S\!N}}\!(z)}
\newcommand{\msnr}{{\rm N_{S\!N}}\!(z,m)}
\newcommand{\snrx}{{\rm N_{S\!N}}\!(x)}
\newcommand{\lsnu}{\mbox{${\cal L}^{\rm S}_\nu$}}
\newcommand{\lsnuv}{\mbox{${\cal L}^{\rm S}_\nu\!(\epsilon^{\prime})$}}
\newcommand{\heat}{{\rm T}_\nu}
\newcommand{\yc}{\mbox{y$_{\rm cut}$}}
\newcommand{\mout}{\langle{\rm M_Z}\rangle}
\newcommand{\nbmout}{{\rm M_Z}}
\newcommand{\scminv}{1/{\rm cm}^2/{\rm s}}
\newcommand{\mevscminv}{1/{\rm cm}^2/{\rm s}/{\rm MeV}}
\newcommand{\aenu}{\langle{\rm E}_\nu\rangle}
\newcommand{\aheat}{\langle\heat\rangle}
\newcommand{\antinue}{\bar{\nu}_{\rm e}}        
\newcommand{\etox}{\nu_{\rm e}\leftrightarrow\nu_{\rm x}}
\newcommand{\mutotau}{\nu_\mu\leftrightarrow\nu_\tau}
\newcommand{\ebtoxb}{\antinue\leftrightarrow\bar{\nu}_{\rm x}}
\def\snii{SN\thinspace{$\scriptstyle{\rm II}$}~}
\def\hi{H\thinspace{$\scriptstyle{\rm I}$}~}

\begin{figure}[b]
\includegraphics[width=13cm]{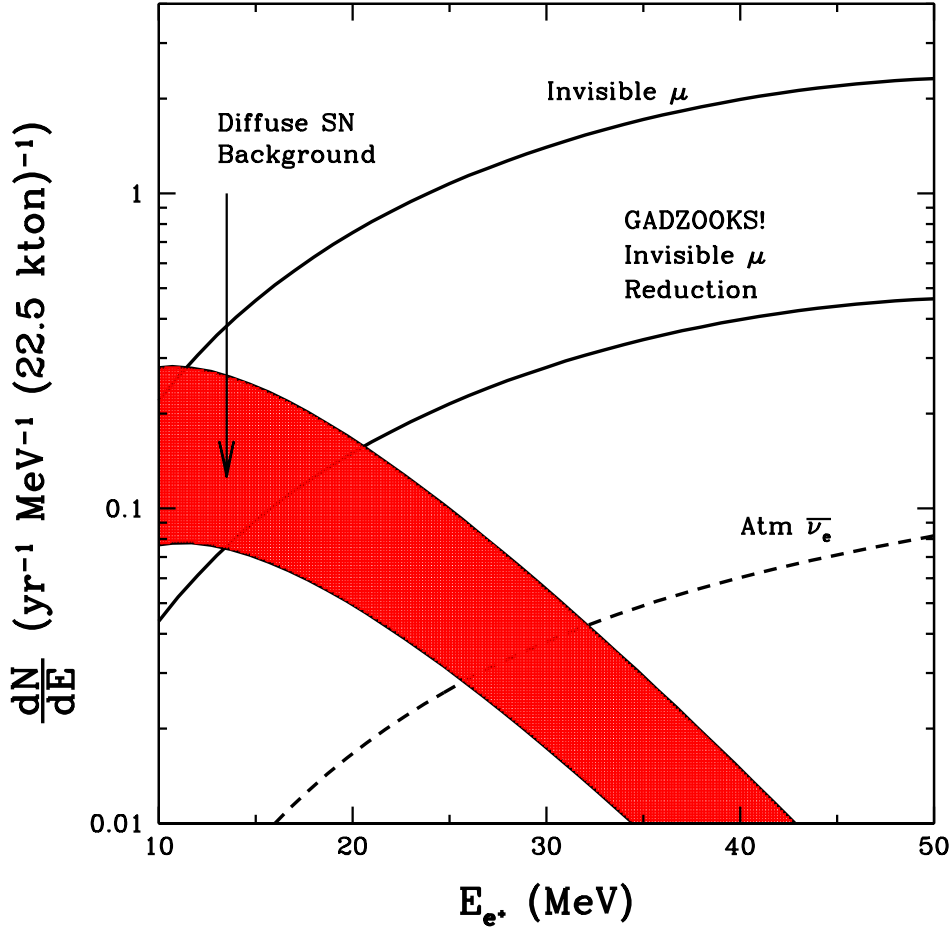}
\caption{ The diffuse supernova neutrino background (shaded) and
competing backgrounds at Super-K and GADZOOKS (a Gd-doped
Super-K). The diffuse SNB is calculated using proxies for the
supernova rate as a function of redshift.  Best estimates for this
rate are just consistent with the Super-K upper limit on the SNB and
correspond to the top of the shaded curve.  The lower border of the
SNB corresponds to the minimum background consistent with estimates of
the supernova rate.
\label{fig:SNrate}}  
\end{figure}

There is a diffuse supernova neutrino background (SNB) coming from all
core collapse supernova (predominantly ``Type II'')
and it is within reach of
current water Cerenkov detectors (\eg, Super-K).  Best
estimates\cite{Strigari} for the expected SNB predict an event rate
that is just consistent with Super-K's upper limit\cite{SK} (see
Fig.~\ref{fig:SNrate}).  The detection of this background provides, at a minimum,
the first detection of cosmological neutrinos (\ie, outside our Local
Group) and an independent measurement of the star formation rate (SFR)
out to a redshift of $z \sim 1$\cite{Fuku}.  For example, the current
Super-K upper-limit to the SNB above 19 MeV ($1.2 {\bar{\nu_e}}{\rm
cm}^{-2} {\rm sec}^{-1}$) already admits constraints on the SFR out to
$z \sim 1$ that are competitive with optical surveys like the Sloan
Digital Sky Survey (SDSS)\cite{Strigari}.  Lower threshold, larger
detectors than Super-K (\eg, a doped Super-K like
GADZOOKS\cite{gadzooks}, KamLAND\cite{kamLAND}, HyperK\cite{hyper}, or
UNO\cite{UNO}) have further reach in redshift and thus can provide
information regarding supernovae and star formation in the
high-redshift Universe, an epoch where photon observations are
complicated by intervening material.

The event rate for ${\bar{\nu_e}}+p\rightarrow n + e^+$ associated
with the diffuse SNB depends on the $\bar{\nu}$ energy spectrum
produced by each core collapse supernova (now set by models, but
possibly to be pinned down by direct direct observations of Galactic
supernovae), an assumed mixing of the $\bar{\nu}$, either in-flight or
within the supernova (constrained by solar, atmospheric, and
variable-baseline neutrino oscillation experiments), and the supernova
redshift distribution.  For Super-K, operating at a threshold of 19
MeV, the neutrino temperatures in the supernova limit the accessible
range to $0\la z \la 1$.  In this range there are several proxies to
the SN rate: the UV and H$\alpha$ luminosity densities, the metal
enrichment rate, and the cosmic optical spectrum as measured by the
SDSS.  Combining these different methods results in a best-estimate to
the SN rate which yields a diffuse SNB that is tantalizingly close the
the Super-K upper bound.  Fig.~\ref{fig:SNrate} shows the allowed range of the
diffuse SNB signal based on the above-mentioned optical proxies as
limited by the Super-K upper limit (star formation rates, acceptable
on the basis of optical surveys, which would produce a diffuse SNB
detectable by Super-K, have been eliminated).  Also shown are the
relevant backgrounds for Super-K and, for comparison, GADZOOKS.  The
lower boundary represents the minimum diffuse SNB consistent with SDSS
observations (similar to other estimates (\cite{Fuku},\cite{ando}) and
representative of the dependence of the SNB on proxies of the SFR).

Because the expected Super-K signal is small, alternative means of
detecting the diffuse SNB should be considered.  Bigger is better
because it allows a more accurate determination of the signal and
therefore more accurate measurements of the high redshift Universe
will result.  A detector which lowers the detection threshold will
automatically increase the redshift sensitivity.  As an example, even
in the relatively small KamLAND detector one has the hope of using the
difference between the Super-K signal and the KamLAND signal
(appropriately normalized) to probe SFRs beyond $z\sim1$.  Roughly 5\%
of Super-K's event rate comes from $z\ga 1$ while roughly 30\% of a
KamLAND detection would be from supernova at $z\ga 1$.  Clearly the
smallness of KamLAND will limit the quality of this comparison and it
would be more desirable to have a low-threshold detector of equal or
greater than Super-K size.


\newpage
\begin{center}
{\Large\sc Measurements of neutrino-nucleus cross sections}\\
\end{center}


Core collapse supernovae are among the most energetic explosions in
our universe, releasing 10$^{46}$ Joules of energy, 98\% in the form of
neutrinos of all flavors which emanate at a staggering rate of 10$^{57}$
neutrinos per second. The energy emitted as visible light, which is
only one ten-thousandth the energy emitted in neutrinos, is enough to
make these explosions as bright as an entire galaxy. These explosions
almost entirely disrupt stars more massive than 8-10 solar masses,
producing and disseminating into the interstellar medium many of the
elements heavier than hydrogen and helium.  Supernovae are a key link
in our chain of origins from the big bang to the formation of life on
earth and serve as laboratories for physics beyond the standard model
and for matter at extremes of density, temperature, and neutron
fraction that cannot be produced in terrestrial laboratories.

As the name suggests, core collapse supernovae result from the
collapse of the core of a massive star at the end of its life. The
collapse proceeds to supernuclear densities, at which point the core
becomes incompressible, rebounds, and launches a shock wave into the
star that is ultimately responsible for the explosion. The shock wave
stalls, however, due to several enervating processes~\cite{bib01}, and
the shock is believed to be revived at least in part by the intense
flux of neutrinos which emanates from the proto-neutron star at the
center of the explosion~\cite{bib02,bib03}. Reactions between this neutrino
flux and the in-falling stellar layers also play a role in the
production of many elements heavier than iron.

Precision neutrino-nucleus cross section measurements are crucial to
improving our understanding of supernovae. Their importance arises in
three related areas: (1) supernova dynamics, (2) supernova
nucleosynthesis, and (3) terrestrial supernova neutrino detection.

(1) Supernova Dynamics

Despite the observational fact that stars do explode, core-collapse
supernova models have historically had difficulty repeating this
feat. One likely culprit is incomplete knowledge of important
micro-physics, in particular, the vitally important contributions of
weak interactions (electron capture and neutrino induced
reactions)~\cite{bib04,bib05}.

Recent studies have demonstrated unequivocally that electron capture
on nuclei plays a major role in dictating the dynamics of stellar core
collapse, which sets the stage for all of the supernova dynamics that
occur after stellar core bounce and the formation of the supernova
shock wave. Past supernova models used naive electron capture rates
based on a simple independent-particle shell model for the nuclei in
the stellar core~\cite{bib06}. Recent supernova calculations use a
model which better captures the realistic shell structure of the
nuclei found in the core and the collective excitations of nucleons in
such nuclei during weak interactions such as electron
capture~\cite{bib07}. Comparisons between these two studies
demonstrate that the more realistic electron capture rates lead to
quantitative and qualitative changes in the stellar core profiles in
density, temperature, and composition after stellar core bounce,
thereby affecting the strength and the launch radius of the supernova
shock wave. These differences have ramifications for both supernova
dynamics and supernova element synthesis.

It is impossible to directly measure weak-interaction electron capture
cross sections at these energies due to the distorting influence of
atomic electrons. Some information relevant to electron capture rates
has been obtained from $(p,n)$ transfer reactions. But these reactions
yield unambiguous information only for the Gamow-Teller part of the
weak operator responsible for electron capture. In addition, at
excitation energies above the Gamow-Teller peak the contributions from
non-zero angular momentum components are difficult, if not impossible
to isolate. Furthermore, even Gamow-Teller peak measurements are
sparse for $A>65$. Measuring cross sections for electron-neutrino
capture on nuclei is equivalent to measuring cross sections for
electron capture on that same nucleus since they are inverse
processes, and it is the only way to make these measurements. Also,
comparisons between $(p,n)$ transfer reaction data and electron-neutrino
capture data for energies above the Gamow-Teller peak will help to
deconvolute the non-zero angular momentum contributions. Thus
experiments to measure neutrino capture rates are complementary to
$(p,n)$ measurements at RIA . This has implications for any application
that requires accurate nuclear structure theory input.

It would, of course, be impossible to experimentally measure cross
sections for all the thousands of weak interaction rates needed in
realistic simulations of supernovae and supernova
nucleosynthesis. Nonetheless, a finite, but strategically chosen set
of measurements will validate the fundamental nuclear structure models
at the foundation of the thousands of rate computations that are input
to the supernova models.

(2) Supernova Nucleosynthesis

Nucleosynthesis in core collapse supernovae falls into three basic
categories: (i) explosive nucleosynthesis that occurs as the shock
wave passes through the stellar layers and causes nuclear fusion
through compression and heating of the material, (ii) neutrino
nucleosynthesis in the ejected layers that occurs as these layers are
exposed to the intense neutrino flux emerging from the proto-neutron
star, which is responsible for generating the explosion to begin with,
and (iii) {\it r}-process nucleosynthesis that occurs in a
neutrino-driven wind emanating from the proto-neutron star after the
explosion is initiated. In all cases, the final elemental abundances
produced and ejected are affected through nuclear transmutations by
the neutrino-nucleus interactions that
occur~\cite{bib08,bib09,bib10,bib11}. Precision neutrino-nucleus cross
section measurements are therefore necessary for a quantitative
understanding. In a supernova many of the reactions take place on
nuclei that unstable and/or are in excited states. Cross sections for
such reactions are obviously not feasible, so the role of neutrino
cross section measurements will be to constrain nuclear structure
theory on a strategically chosen set of nuclei that are similar enough
to the nuclei of interest that reasonable extrapolations can be made.

(3) Supernova Neutrino Detection

An incredible wealth of information was derived from the handful of
neutrinos emanating from supernova SN1987a that were measured in
terrestrial detectors. The time and energy distributions of neutrinos
emanating from a supernova (the ``light curve'' provide the only way
to experimentally explore supernova dynamics at the earliest and
deepest stages of the explosion -- the only way to take a ``picture''
of the birth of a neutron star, and possibly a black hole. The ability
to detect, understand, and ultimately use the detailed neutrino light
curve from a future core collapse supernova in our galaxy is integral
to a) developing better supernova models and b) using precision
supernova models together with detailed astronomical observations in
order to cull fundamental nuclear physics that would otherwise be
inaccessible in terrestrial experiments. In turn, this will require
accurate knowledge of the response function (cross sections and
byproducts of neutrino interactions in the detector material) of a
terrestrial neutrino detector to the incoming supernova neutrino
flux. From deuterium to lead, a number of nuclei have been proposed
and, in some cases, used as supernova detector
materials~\cite{bib12,bib13,bib14,bib15}. In all cases, accurate cross
sections for neutrino-nucleus interactions in the relevant energy
range are essential.

\begin{figure}
\includegraphics[width=8cm]{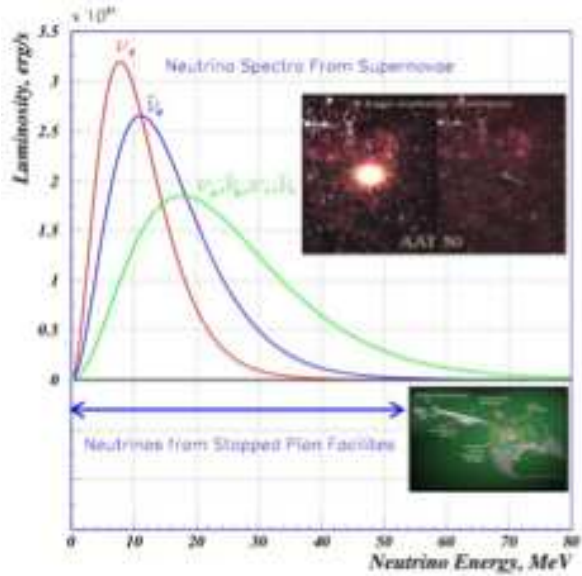}
\caption{The neutrinos produced by stopped-pion facilities are in just
the right energy range to measure the neutrino-nucleus cross sections
required for understanding supernova neutrino emission.
\label{fig:SNS}}
\end{figure}    
 
At this moment there is a unique opportunity to carry out a long term
program of neutrino cross section measurements on a range of
appropriate nuclear targets due to the construction of the Spallation
Neutron Source (SNS) at Oak Ridge National Lab. The SNS generates a
short (380\,ns FWHM), very intense (10$^{14}$ protons/pulse) 60\,Hz, 1\,GeV
proton beam which stops in a mercury target, producing pulsed neutron
beams that will be primarily used for solid-state physics
measurements. As a by-product, very intense neutrino pulses will be
produced through the decay-at-rest (DAR) of pions and muons. This
combination of the DAR neutrino spectra, intensity and time structure
makes the SNS uniquely suited to make an extensive set of precision
neutrino-nucleus cross section measurements:

{\it Energy spectrum}: The neutrino spectrum has strong overlap with the
neutrino spectra present inside a supernova, as shown in Fig.~\ref{fig:SNS}.
Furthermore, the spectrum
is well-defined by the kinematics of the pion and muon decays, making
interpretation of the data much easier.

{\it Intensity}: The SNS will produce 10$^{15}$ neutrinos/sec from
pion and muon decays.  With this flux, a 10\,ton fiducial detector
situated 20 meters from the mercury production target would be
expected to measure several thousand neutrino interactions per year.

{\it Time structure}: The pulsed nature of the beam allows the
detector to be shut off except for a few microseconds after each
pulse, effectively blocking out most cosmic-ray interactions (a
background reduction equivalent to locating the detector underground
with a 2.3\,km water-equivalent overburden).

A suitable location at the SNS (sufficient load-capacity, not
interfering with neutron-scattering instruments, with a volume
sufficient to hold a passive-shielding bunker and an active veto
system plus two 10\,ton fiducial detectors) has been
identified. Detector concepts have been developed which allow
measurement of both solid and liquid targets. In both cases the
detectors can be reused with different target elements, thus
minimizing the costs of a multi-target program. Simulations of these
detectors indicate that there is sufficient angular and energy
resolution of the outgoing electron to allow the double differential
charged current measurements, which is important for using these
measurements to benchmark nuclear structure theory. Neutral current
measurements are relatively difficult because their signal in the
detector is more subtle. However, such measurements may be possible
(given appropriate detector development, and shielding design) during
an SNS proton pulse. During this time the intensity of muon neutrinos,
which result from pion decay, is orders of magnitude higher than the
intensity of other neutrino flavors due to the shorter pion
lifetime. Muon neutrinos also have the benefit of being mono-energetic
since they are a two-body decay product. Neutrino-nucleus interactions
are the only way to get information on neutral current nuclear
excitations.

In addition to the two general-purpose, detectors it would also be
possible to install elements of supernova neutrino detectors in the
shielded bunker and {\it directly} calibrate them with a known flux of
neutrinos (flavor and energy) having an energy spectrum overlapping
that of supernova neutrinos.

A broad range of precision neutrino nucleus interaction measurements
are key to understanding supernovae -- both nucleosynthesis and the
explosion mechanism -- and are therefore key to understanding our
star dust origins. Such measurements also provide input to nuclear
structure theory that is unique and complementary to that which will
be obtained at RIA. Constructing the ideal neutrino source with which
to make such measurements would require an enormous
investment. However, this ideal source -- the Spallation Neutron
Source -- is already under construction for completely different
purposes. The neutrinos which are produced are a serendipitous
by-product, providing us with a unique opportunity to make these
important measurements.


\newpage
\begin{center}
{\Large\sc Leptogenesis and the origin of the baryon asymmetry}\\
\end{center}


One of the most striking impact of neutrino physics on cosmology is a
possible explanation to the baryon asymmetry of the universe, called
``leptogenesis.''  This is a surprising connection, as the {\it
  baryon}\/ asymmetry as we see from our own existence and in the Big
Bang Nucleosynthesis is in {\it quarks}\/, not in neutrinos.

\begin{figure}
  \centering
  \includegraphics[width=0.25\textwidth]{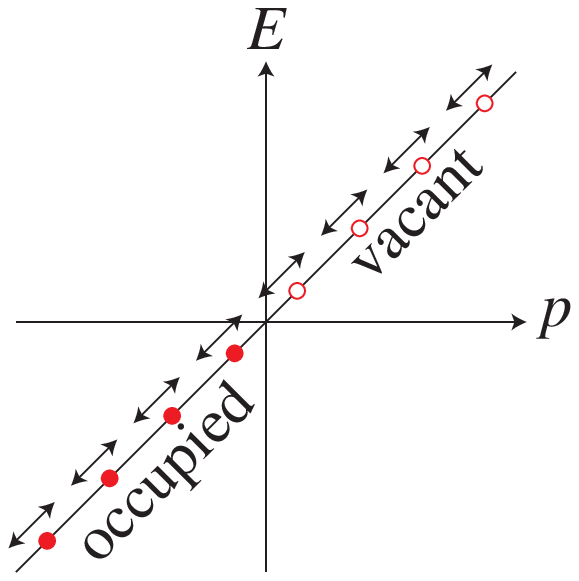}\hspace{0.15\textwidth}
  \includegraphics[width=0.25\textwidth]{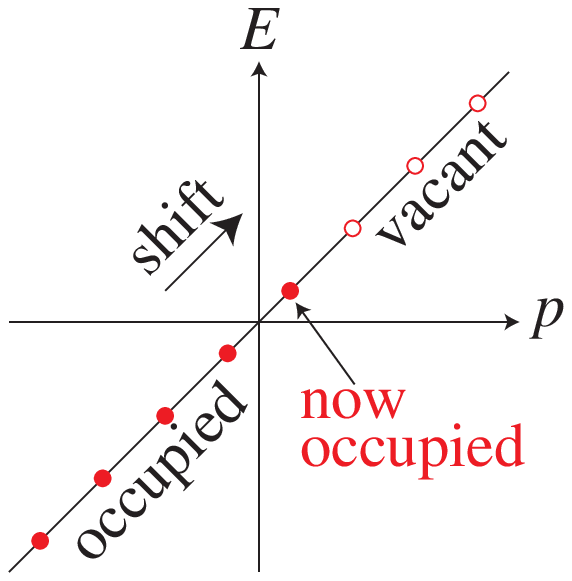}
  \caption{Schematic explanation of the electroweak anomaly effect.
    (1) Negative energy states of the Dirac equation are occupied
    while positive energy states are left vacant.  Thermal fluctuations
    of the $W$ field move the energy levels up and down.  (2) Once in
    a while, the fluctuation grows so big that all energy levels are
    shifted by one unit.  Then a particle is found occupying a
    positive energy state.  The same process occurs for all particles
    coupled to the $W$-boson, and hence left-handed leptons and
    left-handed quarks of all three colors change their numbers by the
    same amount; hence $\Delta L = \Delta B$. \label{fig:schematic}}
\end{figure}

The crucial ingredient in the connection between neutrino physics and
baryon asymmetry is the electroweak anomaly, an effect in the {\it
  Standard Model}\/.  As first pointed out by 't Hooft
\cite{'tHooft:up}, the baryon number $B$ is conserved classically in
the Standard Model, but not quantum mechanically.  Similarly, the
lepton number $L$ is not conserved either, while $B-L$ is.  The
anomaly is a tunneling effect at the zero temperature and hence highly
suppressed by a WKB-like factor (``instanton'') $e^{-2\pi/\alpha_W}
\approx 10^{-1517}$.  However, at high temperatures too hot for the
Higgs Bose-Einstein condensate, it is not a tunneling because
classical transitions are allowed by the thermal fluctuation of
electroweak gauge fields in the plasma \cite{Kuzmin:1985mm}.  Then the
rate of anomaly-induced $B$- and $L$-violating process is suppressed
only by $10 \alpha_W^5 \approx 10^{-6}$ \cite{Moore:2000mx}.  In fact,
$B$- and $L$-violation is in equilibrium for the range of temperatures
approximately for $10^2$--$10^{11}$~GeV.  This anomaly effect can be
understood schematically in Fig.~\ref{fig:schematic}.

If, due to some reason, there is net asymmetry in $B-L$, the chemical
equilibrium in gauge and Yukawa interactions as well as the anomaly
effects is obtained at \cite{Harvey:1990qw}
\begin{equation}
  \label{eq:1}
  B \simeq 0.35 (B-L), \qquad L \simeq -0.65 (B-L).
\end{equation}
The important question is then how an asymmetry in $B-L$ had been
created. It obviously requires violation of $B-L$.  The most popular
mechanism for it is the seesaw mechanism \cite{seesaw}.  It postulates
heavy right-handed neutrinos $N$ with no gauge
interactions (they may well have gauge interactions at yet
higher energy scales, such as in $SO(10)$ grand unified theories)
but with Yukawa interactions
\begin{equation}
  \label{eq:2}
  {\cal L} = - \left(\frac{M_\alpha}{2} N_\alpha N_\alpha
    + h_{\alpha i} N_\alpha L_i H + c.c.\right) .
\end{equation}

One can always choose the basis for $N_\alpha$ such that
their masses are diagonal and real positive $M_\alpha > 0$, while the
Yukawa couplings $h_{\alpha i}$ are in general three-by-three complex
matrix with 18 independent parameters.  Using the rephasing of charged
lepton fields, the physical parameters in $h_{\alpha i}$ are reduced
to 15.  On the other hand, CP is conserved only if there is a basis
where all entries of $h_{\alpha i}$ are real, and hence with only 9
parameters.  Therefore there are $15-9=6$ CP-violating phases in
general.  Only one of them can appear in neutrino oscillation, two in
Majorana phases, and three others appear only in processes that
involve the right-handed neutrinos directly.

\begin{figure}
  \centering
  \includegraphics[width=0.7\textwidth]{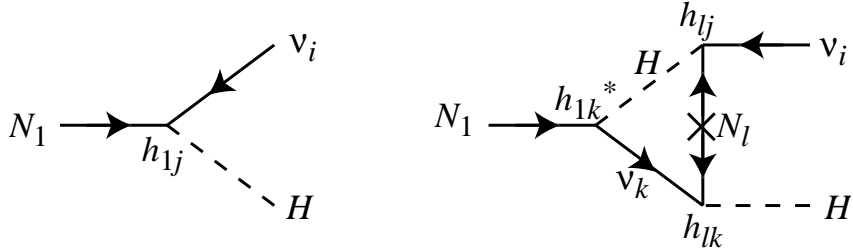}
  \caption{The interference between the tree-level and the one-loop
  amplitudes causes the decay asymmetry.\label{fig:interference}} 
\end{figure}

In early universe, $N_\alpha$ were present in the thermal bath, and
decayed eventually.  If $\sum_i |h_{\alpha i}|^2 \lesssim
M_\alpha/(10^{16}~{\rm GeV})$, they have too small decay and inverse
decay rates to stay in thermal equilibrium.  They ``hang around'' for
a while before they decay at a temperature much below their masses.
On the other hand, the decay may be CP violating.  At the tree-level,
$N_\alpha$ decays equally into $L H$ and its CP conjugate.  At the
one-loop level, however, the interference between the tree-level
amplitude and the absorptive part in the one-loop level amplitude
(Fig.~\ref{fig:interference}) results in a direct CP asymmetry
\cite{Fukugita:1986hr}
\begin{equation}
  \label{eq:3}
  \epsilon_1 \equiv BR(N_1 \rightarrow LH) - BR(N_1 \rightarrow
  \overline{(LH)}) 
  \simeq \frac{1}{2\pi} \frac{\sum_{i,j}\Im(h_{1i} h_{1j} h_{3i}^* h_{3j}^*)}
  {\sum_i |h_{1i}|^2} \frac{M_1}{M_3}.
\end{equation}
Here, we assumed that the relevant decay is that of $N_1$, ignored
loops in $N_2$, and assumed a hierarchical spectrum $M_3 \gg M_1$.
One can show that this quantity cannot exceed \cite{Davidson:2002qv}
\begin{equation}
  \label{eq:4}
  |\epsilon_1| \leq \frac{3}{16\pi} \frac{M_1}{v^2} \sqrt{\Delta
    m^2_{23}} \simeq 10^{-6} \left(\frac{M_1}{10^{10}~{\rm GeV}}\right)
  \left(\frac{m_3}{0.05~{\rm eV}}\right).
\end{equation}
To produce a sufficient baryon asymmetry, one finds a lower limit on
the right-handed neutrino mass of $4 \times 10^{8}$~GeV, and
correspondingly an upper limit on the neutrino mass $m_i < 0.12$~eV
\cite{Buchmuller:2004nz}.  These bounds, however, can be
  evaded if there is an extreme mass degeneracy among right-handed
  neutrinos within their widths because the indirect CP violation in
  their mixing can enhance the resulting asymmetry.

Because right-handed neutrinos are required to be heavy, the reheating
temperature after the inflation must be high, $T_{RH} > 3 \times
10^9$~GeV \cite{Buchmuller:2004nz}.  This high temperature is in
conflict with the gravitino problem if there is supersymmetry.  For
$m_{3/2} \simeq 100$~GeV--1~TeV, and for purely photonic decay
$\tilde{G} \rightarrow \tilde{\chi} \gamma$, the reheating temperature
must be below $10^6$--$10^9$~GeV \cite{Kawasaki:1994af}.  For hadronic
decay, the bound may be even tighter \cite{Kawasaki:2004yh}.  There
are several ways out of this conflict.  One is to assume very heavy
gravitino $m_{3/2} \gtrsim 50$~TeV as in anomaly-mediated
supersymmetry breaking \cite{Randall:1998uk,Giudice:1998xp}; then the
gravitinos decay before the Big-Bang Nucleosynthesis and are harmless.
The reheating temperature can be as high as $10^{11}$~GeV in this case
\cite{Kawasaki:1994af}, sufficient for thermal leptogenesis.  Another
possibility is to assume non-thermal production of right-handed
neutrinos.  They may be produced directly by inflaton decay
\cite{Lazarides:wu}, or their scalar partners may be the source of the
energy density \cite{Murayama:1993em}.  Right-handed sneutrino may
even be the inflaton itself \cite{Murayama:1992ua}.  Then the
reheating temperature can be as low as $10^6$~GeV to suppress
gravitino production, while a sufficient baryon asymmetry can be
obtained \cite{Hamaguchi:2001gw}.

Finally, it is important to note that the Majorana-ness of neutrinos is
not mandatory for leptogenesis \cite{Dick:1999je,Murayama:2002je}.  If
neutrinos are Dirac, there is no lepton number violation.  On the
other hand, the right-handed neutrinos are light (degenerate with
left-handed ones by definition) and have only Yukawa interaction as
small as $10^{-13}$.  Therefore, even if there is no overall lepton
asymmetry, it is possible to ``store'' asymmetry in right-handed
neutrinos while the asymmetry in the other leptons is shared with
quarks via the anomaly.

Can we prove leptogenesis experimentally?  Clearly we need CP
violation in the neutrino sector for leptogenesis.  Unfortunately, the
CP violation we can probe in neutrino oscillation may not be the CP
violation needed in leptogenesis.  In some less general models of
neutrino mass, they are correlated ({\it e.g.}\/, the model with only
two $N$ \cite{Frampton:2002qc}).  Nonetheless the observation of CP
violation and neutrinoless double beta decay would provide strong
circumstantial evidence for leptogenesis.  Further model-dependent but
supporting evidence may be found in lepton-flavor violation if
right-handed neutrino interactions leave imprint in mass matrices of
sleptons in supersymmetric models (see, {\it e.g.}\/,
\cite{Ellis:2002fe,Davidson:vc,Ibe:2004tg}).  Even though we may not
be able to convict leptogenesis in a criminal trial, we may still find
it guilty in a civil case.


\newpage
\begin{center}
{\Large\sc Precision big bang nucleosynthesis tests}\\
\end{center}

\def\he#1{\iso{He}{#1}}
\def\li#1{\iso{Li}{#1}}
\def\iso#1#2{\mbox{${}^{#2}{\rm #1}$}}
\newcommand\nnu{N_{\rm \nu,eff}}
\newcommand\beq{\begin{equation}}
\newcommand\eeq{\end{equation}}
\newcommand\beqar{\begin{eqnarray}}
\newcommand\eeqar{\end{eqnarray}}

Big bang nucleosynthesis is the cosmological theory of the origin of 
the light element isotopes D, $^3$He, $^4$He, and $^7$Li \cite{bbn}.
The success of the theory when compared to the observational
determinations of the light elements allows one to place strong 
constraints on the physics of the early Universe at a time scale of 
1-100 seconds after the big bang.  $^4$He is a sensitive probe 
of deviations from the standard model and its abundance is determined 
primarily by the neutron to proton ratio when nucleosynthesis begins 
at a temperature of $\sim 100$ keV (to a good approximation all 
neutrons are then bound to form $^4$He).

\begin{figure}[b]
\includegraphics[height=14cm]{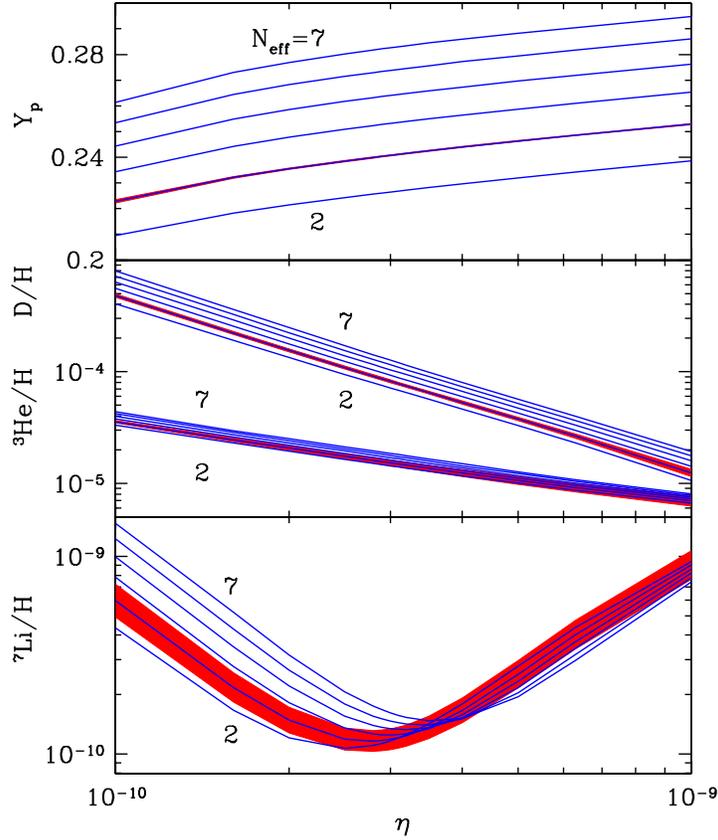}
\caption{The light element abundances as a function of the
baryon-to-photon ratio for different values of $N_\nu$
\protect\cite{cfo2}.
\label{fig:3}}
\end{figure}

The ratio $n/p$ is determined by the competition between the weak
interaction rates which interconvert neutrons and protons,
\begin{equation}
 p + e^- \leftrightarrow n + \nu_e \, , \qquad 
 n + e^+ \leftrightarrow p + \bar \nu_e \, , \qquad 
 n \leftrightarrow p + e^- + \bar \nu_e
\label{weakrates}
\end{equation}
and the expansion rate, and is largely given by the Boltzmann 
factor
\begin{equation}
  n/p \sim e^{-(m_n - m_p)/T_f}
\label{nperp}
\end{equation}
where $m_n-m_p$ is the neutron to proton mass difference.  The weak
interactions freeze out at a temperature of roughly 1 MeV when the
weak interaction rate, $\Gamma_{\rm wk} \sim {G_F}^2 {T}^5$ is
comparable to the Hubble expansion rate, $H(T) \sim \sqrt{G_N N}
{T}^2$, where $N = g_\gamma + {7 \over 8} g_e + {7 \over 8} g_\nu
N_\nu = {11\over 2} + {7 \over 4}N_\nu$ for $g_\gamma = g_\nu = 2$ and
$g_e = 4$ and $N_\nu$ is the number of neutrino flavors.  Freeze-out
is then determined by
\begin{equation}
{G_F}^2 {T_f}^5 \sim  \sqrt{G_N N} {T_f}^2
\label{comp} \label{freeze}
\end{equation}
The freeze-out condition implies the scaling $T_f^3 \sim \sqrt{N}$.
From Eqs.~(\ref{nperp}) and (\ref{freeze}), it is then clear that
changes in $N$, caused for example by a change in the number of light
neutrinos $N_\nu$, would directly influence $n/p$, and hence the
$^4$He abundance, which is given by $Y_p = 2(n/p)/[1+(n/p)]$.  The
dependence of the light element abundances on $N_\nu$ is shown in
Figure~\ref{fig:3} \cite{cfo2}, where plotted is the mass fraction of
$^4$He, $Y$, and the abundances by number of the D, $^3$He, and $^7$Li
as a function of the baryon-to-photon ratio, $\eta$, for values of
$N_\nu = 2 - 7$. As one can see, an upper limit to $Y$, combined with
a lower limit to $\eta$ will yield an upper limit to $N_\nu$
\cite{ssg}.  It should be noted that although the number of light
neutrino flavors has been fixed by experiments at LEP and SLAC, the
BBN bound is not solely restricted to neutrinos, but rather any
relativistic particle species present in the early Universe at the
time of BBN.  In this sense, $N_\nu > 3$ is simply a surrogate for any
new particle species.

\begin{figure}[b]
\includegraphics[height=13cm]{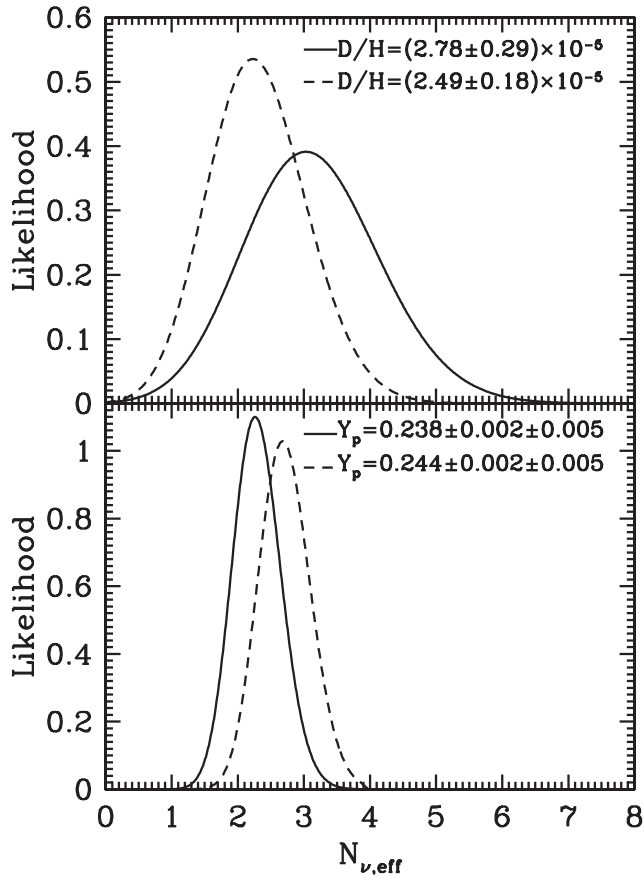}
\caption{\label{fig:N_nu}
Likelihoods for $\nnu$ as predicted by the WMAP 
$\eta$ and light element observations:
{\bf (a)} deuterium 
{\bf (b)} helium.}
\end{figure}
 
Assuming no new physics at low energies, the value of $\eta$ is the
sole input parameter to BBN calculations. Historically, it has been
fixed by the comparison between BBN predictions and the observational
determinations of the isotopic abundances. However, the high precision
results from WMAP \cite{wmap}, have determined the primordial spectrum
of density fluctuations down to small angular scales with excellent
agreement with galaxy and cluster surveys and these results have led
to a determination of the baryon density of unprecedented accuracy.
The WMAP result of $\Omega_B h^2 = 0.0224 \pm 0.0009$ is equivalent to
$\eta_{\rm 10} = 6.14 \pm 0.25$, where $\eta_{10} = 10^{10}
\eta$. This result is the WMAP best fit assuming a varying spectral
index and is sensitive mostly to WMAP alone (primarily the first and
second acoustic peaks) but does include CBI \cite{cbi} and ACBAR
\cite{acb} data on smaller angular scales, and Lyman $\alpha$ forest
data (and 2dF redshift survey data \cite{2df}) on large angular
scales.

With the value of $\eta$ fixed, one can use He abundance measurements
to set limits on new physics \cite{sarkar}.  At the WMAP value of
$\eta$, the \he4 abundance is predicted to be $Y_P =
0.2484^{+0.0004}_{-0.0005}$ \cite{cfo3} and is somewhat high compared
with observationally determined values of $Y_P = 0.238 \pm 0.002 \pm
0.005$ \cite{yp} or $Y_P = 0.242 \pm 0.002 \pm 0.005$ \cite{it}.  This
discrepancy likely is due to systematic errors (possibly due to
underlying He absorption \cite{os}).  Indeed a preliminary analysis
with these effects included shows much better agreement \cite{os2}.
Until this situation is better understood, caution is in order.

The $\nnu$ likelihood calculated \cite{cfo2,cfo3,fior} using observed
\he4 abundances appears in Fig.~\ref{fig:N_nu}b.  As pointed out
above, all available \he4 abundance observations fall short of the
CMB-BBN predicted value.  This shortfall manifests itself in
Fig.~\ref{fig:N_nu}b by driving $\nnu$ down below 3 for both observed
\he4 abundances, to $\nnu \approx 2.5$.  The width of these
distributions is quite narrow, $\Delta\nnu \approx 0.4$, due to the
strong sensitivity of \he4 to $\nnu$.  Indeed, the width of the
likelihood is dominated by the large systematic uncertainties in the
\he4 observations.  In order for this constraint to be considered
robust, we must understand the hidden systematics in the \he4
observations.  Assuming a prior of $\nnu \ge 3.0$ the corresponding
95\% CL upper limits are: $\nnu < 3.4$ for $Y_P = 0.238$; $\nnu < 3.6$
for $Y_P = 0.244$. When underlying absorption is included, the upper
limit on $\nnu$, may be significantly increased.

On the other hand, deuterium may not appear to suffer from large
systematics.  It is, however, limited by the low number statistics due
to the difficulty of finding high-redshift systems well-suited for
accurate D/H determinations.  Given that D predictions from WMAP agree
quite well with observations, we can now use D to place an interesting
limit on $\nnu$ \cite{cfo3}.  D is not as sensitive to $\nnu$ as \he4
is, but none-the-less it does have a significant dependence.  The
relative error in the observed abundance of D/H ranges from 7-10\%,
depending on what systems are chosen for averaging.  If the five most
reliable systems are chosen, the peak of the $\nnu$ likelihood
distribution lies at $\nnu \approx 3.0$, with a width of $\Delta\nnu
\approx 1.0$ as seen in Fig.~\ref{fig:N_nu}a.  However, if we limit
our sample to the two D systems that have had multiple absorption
features observed, then the peak shifts to $\nnu \approx 2.2$, with a
width of $\Delta\nnu \approx 0.7$.  Given the low number of
observations, it is difficult to qualify these results.  The
differences could be statistical in nature, or could be hinting at
some underlying systematic affecting these systems.  Adopting the five
system D average, D/H$ = (2.78 \pm 0.29)\times 10^{-5}$, the upper
limit on $\nnu$ is $\nnu < 5.2$, assuming the prior $\nnu \ge 3$.


\newpage
\begin{center}
{\Large\sc Precision cosmic microwave background tests}\\
\end{center}


The WMAP experiment showed that the standard cosmological 
model is a good phenomenological description of the observed 
universe \cite{bennet03a}. In combination with other observations 
(supernova Ia and large scale structure), the consistent picture 
that emerges has matter contributing about 30\% and dark energy 
about 70\% to the energy density of the universe. Most of the 
matter is dark; baryons contribute about 4\% while neutrinos 
contribute less than 2\% to the total energy density of the 
universe. Given that the basic phenomenological structure is 
in place, one can look forward to the future with some confidence. 
The CMB has much more to offer if smaller scales and smaller 
features can be probed. This will require high angular resolution, 
high sensitivity experiments. 

There are three questions of relevance for neutrino physics 
that CMB (and in general cosmology) could help answer. 
How many? What are their masses? How do they get massive? 
The issue of ``how many'' is related to the number of light 
degrees of freedom which is sensitive to the presence of 
sterile neutrinos. It should be noted that if the apparent LSND 
excess of $\bar \nu_e$ is not due to systematic errors then a 
sterile (fourth) neutrino is required. It is possible for the 
abundance of this sterile neutrino to be cosmologically 
relevant. Also, depending on the mass hierarchy, the 
required (LSND) mass-squared difference could affect the overall 
mass scale. The MiniBooNE experiment currently underway at 
Fermilab will soon test the LSND neutrino oscillation hypothesis. 

The fact that the CMB could have imprints of 
the mass generation mechanism might seem surprising, but a 
couple of examples should clarify this connection. Certain
neutrino mass terms permit the possibility of a heavy
neutrino decaying into a lighter one and a scalar particle
(called Majoron). Long lifetime decays of this kind can be
constrained using the CMB \cite{kaplinghat99}. A more 
interesting example is that of neutrino mass generation 
through spontaneous breaking of approximate lepton flavor 
symmetries at or below the weak scale \cite{chacko03}. 
The presence of light Pseudo-Goldstone bosons changes the 
CMB anisotropy through the introduction of new degrees 
of freedom and new scattering channels for the 
neutrino \cite{chacko03}. 

The current large scale structure surveys (2dFGRS, SDSS) and 
WMAP together already provide powerful constraints on neutrino 
mass. We know that the sum of the active neutrino masses is 
less than about 1 eV \cite{spergel03}. The sum of the active 
neutrino masses, $m_\nu$, is related to their energy density 
$\rho_\nu$ as $m_\nu \approx \rho_\nu/({\rm meV})^4$. 
At the lower end, atmospheric neutrino oscillations constrain
the mass of at least one active neutrino to be larger than about 
0.05 eV. It is indeed wonderful that this window from 0.05 eV 
to 1 eV can be probed with both laboratory experiments and 
cosmological observations. 

Another parameter relevant for neutrino physics that future 
CMB experiments can measure well is the number of light degrees 
of freedom $N_\nu$, traditionally labeled ``number of neutrinos''. 
It measures the energy density of relativistic particles in 
units of the energy density of one active neutrino species. 
Significant improvement in the measurement of both $m_\nu$ 
and $N_\nu$ (using CMB) will require future precision 
measurement of the CMB anisotropy from few to 20 arcminute 
angular scales. 

Changing $N_\nu$ affects the CMB anisotropy imprinted on the 
last scattering surface (called the primary CMB). There are 
two predominant effects. First, changing $N_\nu$ changes the 
expansion rate of the universe. At last scattering, this leads 
to a change in the sound horizon and damping length (of the 
photon-baryon fluid). The change in the sound 
horizon shifts the position of the peaks and troughs in the 
anisotropy spectrum while the change in the damping length 
(relative to the sound horizon) changes its amplitude. The 
second effect operates around the location of the first peak 
and on angular scales larger than that. The presence (or lack) 
of radiation has an effect on the CMB even after last scattering. 
The reason is that in the presence of radiation (in fact 
anything other than pressureless matter) the gravitational 
potential changes with time (decays). The photons traversing 
these potential wells pick up a net red-shift or blue-shift 
which enhances the amplitude of the anisotropy spectrum. 

A change in $m_\nu$ gives rise to all of the above effects (though
the changes are not degenerate with that of $N_\nu$). However, a 
massive neutrino has an additional effect. On small scales, the 
presence of a massive neutrino damps the growth of structure 
(gravitational potential or equivalently the matter 
density perturbations). This is simply due to the larger thermal 
speed of the neutrino (as compared to that of dark matter). On 
large enough scales, the neutrino behaves like dark matter while 
on small scales, it moves freely in and out of dark matter 
potential wells. This effect can be used to put constraints
on the neutrino mass using the observed galaxy power spectrum 
combined with CMB observations \cite{hu98a}. Eisenstein et al. 
\cite{eisenstein99} found that just the primary CMB spectrum from 
the Planck satellite can measure neutrino mass with an error of 
0.26 eV. This sensitivity limit is related to the temperature of 
the photons at last scattering, 0.3 eV. No significant improvement is 
expected from combining Planck and the SDSS galaxy power 
spectrum \cite{hu98,eisenstein99}.

\begin{figure}
\includegraphics[width=14cm]{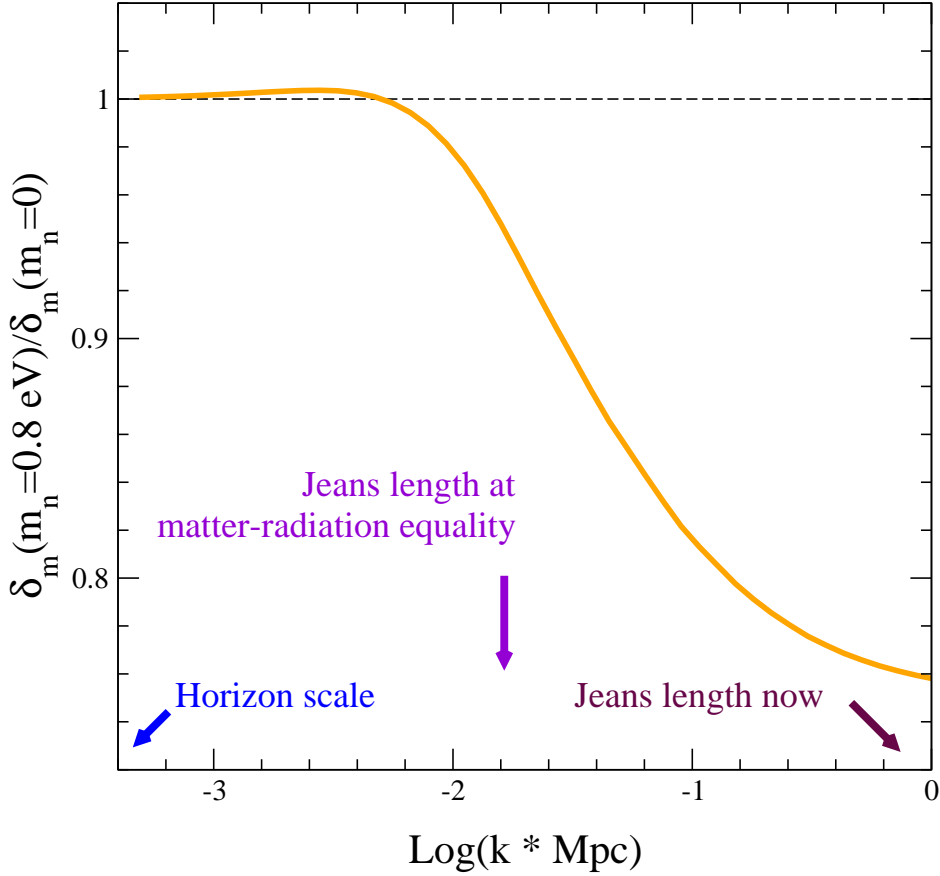}
\caption{The change in the matter density perturbation 
spectrum (as a function of inverse scale $k$) when one 
neutrino is given a mass of 0.8 eV.
\label{fig:tf}}
\end{figure}

The alteration of the gravitational potential at late times 
changes the gravitational lensing of CMB photons as they 
traverse these potentials \cite{seljak96,bernardeau97}. Lensing 
results in the deflection of the CMB photons. These deflections 
are small, on average of the order of arcminutes. However, since 
the structure giving rise to these deflections is 
correlated on large scales, the lensing deflections are 
correlated over large angular scales. Including the 
gravitational lensing effect, the Planck error forecast 
improves to about 0.15 eV \cite{kaplinghat03}.  
More ambitious CMB experiments could reduce
this error to $\sim 0.05$ eV \cite{kaplinghat03}.  
Tomographic observations of  the galaxy shear due to gravitational 
lensing can a achieve similar sensitivity in $m_\nu$ 
\cite{abazajian02}. The physics in both cases is the same:
gravitational lensing. However the observations and the associated
systematics are very different. Complementary techniques are 
valuable since these measurements will be very challenging.

The effect of lensing on the CMB is simple to write down. For example,
the {\em lensed} temperature $T_{\rm U}$ in a particular direction 
$\vec \theta$ is related to the {\em unlensed} temperature 
$T_{\rm L}$ by the relation 
$T_{\rm L}(\vec \theta) = T_{\rm U}(\vec \theta + \vec d)$. The 
lensing deflection field $\vec d$ is related to the underlying 
density perturbations (gravitational potential wells). Thus
a measurement of the statistical properties of this deflection 
field can in principle be translated into a measurement of the 
underlying density perturbations.

Lensing has three main effects on the CMB anisotropy spectrum. 
First, lensing introduces non-gaussianity (of a specific kind) 
into the otherwise gaussian CMB sky maps. This can be used to 
estimate the deflection field from CMB sky maps \cite{hu02}. 
Second, lensing generates a specific pattern of polarization 
called B mode polarization (where B is used in the sense 
of ``gradient--free''), the contribution from which is 
otherwise expected to be small \cite{zaldarriaga98b}. If we are 
to use lensing to measure neutrino mass, this effect will play 
a vital role. Third, lensing due to its intrinsic non-gaussian 
nature shifts power from one scale to another. The net effect 
for the CMB anisotropy spectrum is that lensing becomes an 
important source of power on small scales because the primary 
(unlensed) CMB has an exponential drop in power on small scales.

The signature of a (say) 0.1 eV neutrino in the primary 
(unlensed) CMB anisotropy spectra is small. Such small 
masses are only detectable through their effect on lensing, 
which comes through their influence on the gravitational 
potential. The net suppression of the power spectrum is 
scale dependent and the relevant length scale is the 
Jeans length for neutrinos \cite{bond83,ma96,hu98} which
decreases with time as the neutrino thermal speed decreases.
This suppression of growth is ameliorated on scales larger than 
the Jeans length at matter--radiation equality, where the 
neutrinos can cluster. Neutrinos never cluster on scales 
smaller than the Jeans length today. The net result is no 
effect on large scales and a suppression of power on small 
scales. This explains the scale dependence of density 
perturbations in the presence of a massive neutrino as
plotted in Figure \ref{fig:tf}.  

Future experiments like Planck will be able to statistically 
detect the lensing effect and thus measure or put upper limits on 
the neutrino mass. The expected 1-$\sigma$ error on $m_\nu$ 
from Planck is 0.15 eV while that on $N_\nu$ is 0.2. Planck maps
combined with that from a ground--based but more sensitive 
experiment like the South Pole Telescope could do even better. 
Assuming the South Pole Telescope will run with polarized 
detectors, one could get up to 30\% improvement in the 
expected errors.

Looking beyond Planck, it is conceivable that there will be another 
full sky mission. The primary aim of such an experiment will be to 
measure the primordial (inflationary) B mode signal which is 
expected to be present on large 
scales \cite{kamionkowski97,seljak97}. The sensitivity and 
angular resolution required to measure the primordial B 
mode signal \cite{knox02,kesden02} will allow one 
to achieve a lot more. If the foregrounds can be tamed, then one
could hope to achieve a 1-$\sigma$ error of 0.05 eV on $m_\nu$ 
and 0.1 on $N_\nu$. This would be spectacular.


\newpage
\begin{center}
{\Large\sc Neutrino mass and large scale structure}\\
\end{center}

\newcommand\apjl{Astrophys. J. Lett.}
\newcommand\mnras{Mon. Not. Roy. Astron. Soc.}
\newcommand{\Rf}[1]{\ref{fig:#1}}
\newcommand{\rf}[1]{\ref{fig:#1}}
\def\mtwo{m_{200}}
\def\ang{\,{\rm\AA}}

\def\be{\begin{equation}}
\def\ee{\end{equation}}
\def\bea{\begin{eqnarray}}
\def\eea{\end{eqnarray}}
\newcommand{\vs}{\nonumber\\*}
\newcommand{\ec}[1]{Eq.~(\ref{eq:#1})}
\newcommand{\Ec}[1]{(\ref{eq:#1})}
\newcommand{\eql}[1]{\label{eq:#1}}
\newcommand{\dcl}{\chi_L}
\newcommand{\dcs}{\chi_S}
\newcommand{\rhos}{\rho_s}
\newcommand{\rs}{r_s}
\def\fun#1#2{\lower3.6pt\vbox{\baselineskip0pt\lineskip.9pt
  \ialign{$\mathsurround=0pt#1\hfil##\hfil$\crcr#2\crcr\sim\crcr}}}
\def\la{\mathrel{\mathpalette\fun <}}
\def\ga{\mathrel{\mathpalette\fun >}}


Structure in the universe forms differently if neutrinos have non-zero
mass. In a universe without massive neutrinos, all matter (all massive
particles) participates in the gravitational collapse that begins when
the universe is about one hundred thousand years old. If neutrinos
have mass, they constitute some of the matter today, but they were
very hot early on, so they did not participate in collapse until they
cooled sufficiently.  Therefore, matter in a universe with massive
neutrinos is more clustered than matter in a universe with massless
neutrinos. It is this simple principle, well-known for over twenty
years~\cite{bond}, that leads to the most stringent cosmological
constraint on neutrino masses.

This constraint is based on the textbook calculation\cite{mc} that
there are $112$ neutrinos cm$^{-3}$ for each generation. From this
predicted number density, which follows directly from the
thermodynamics of the universe at temperatures of order one MeV, we
infer that the energy density of massive neutrinos compared to the
critical density is
\be
\Omega_\nu = 0.02 \left( {\sum m_\nu \over 1 eV}\right) 
\left( {72\ {\rm km\ sec}^{-1} {\rm Mpc}^{-1}\over H_0} \right)^2
\equiv 0.02 \left( {\sum m_\nu \over 1 eV}\right) 
\left( {0.72\over h} \right)^2
\ee
where the sum is over the three generations and $H_0$ is the Hubble
constant, known to better than $10\% $. The statistical mechanics of
the early universe also dictates that cosmic neutrinos have the
thermal distribution of a massless gas (occupation number
$[e^{p/T_\nu}+1]^{-1}$) with temperature $T_\nu=(4/11)^{1/3}T_{\rm
cmb}$. A massive neutrino therefore has a thermal velocity of order
$T_\nu/m_\nu \sim 2\times 10^{-4} (1+z) (1 \ {\rm eV}/m_\nu)$ where
$z$ is the cosmic redshift. These thermal velocities enable neutrinos
to freestream out of perturbation regions smaller than $\lambda_{\rm fs}
\simeq 1\, {\rm Mpc}\, (1 \ {\rm eV}/m_\nu) (1+z)^{1/2}$. On length
scales smaller than one Mpc, less matter is available to form potential
wells, so perturbations have been suppressed for all times by the
inability of neutrinos to cluster. There is thus a constant
suppression in the power spectrum (which measures the clumpiness of
the universe) on scales $k\ga 1$ Mpc$^{-1}$, as shown in
Fig.~\ref{fig:tf}
On slightly larger scales, neutrinos can participate in
gravitational collapse when the freestreaming scale becomes smaller
than the scale in question, so the suppression is not as severe. As
Fig~\ref{fig:tf} illustrates, there is thus a monotonic decrease
in the clustering strength as one moves from the largest scale
affected (the horizon when perturbations begin to grow at matter
domination) to the freestreaming scale today. This monotonic
suppression from $k\sim 10^{-2}$ Mpc$^{-1}$ to $k\sim 1$ Mpc$^{-1}$ is
a unique signature of massive neutrinos. The amplitude of the
suppression depends only on the ratio of the massive neutrino density
to the total matter density, $f_\nu\equiv \Omega_\nu/\Omega_m$.

\begin{figure}
\includegraphics[width=14cm]{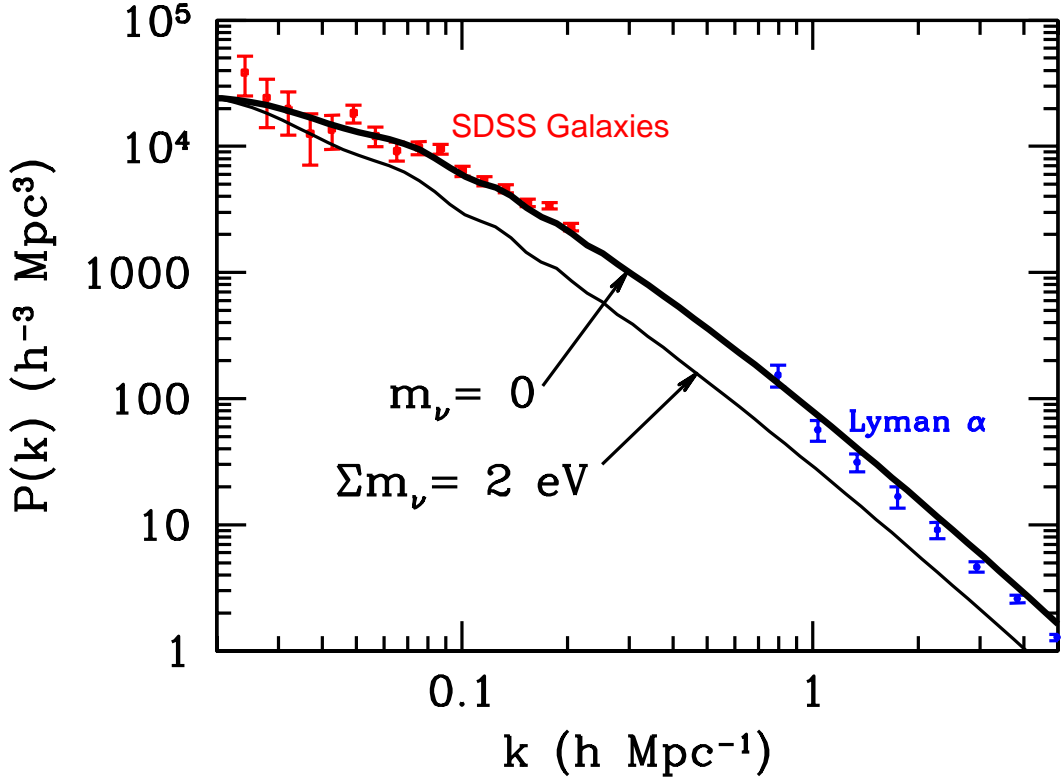}
\caption{The power spectrum of the matter distribution. A non-zero
neutrino mass suppresses power on small scales (large $k$).  Data are
from the Sloan Digital Sky Survey galaxies~\cite{tegmark} and the
Lyman alpha forest~\cite{croft}. There is significant normalization
uncertainty in both data sets and in the theory curves.
\label{fig:scott}}
\end{figure}

The power spectrum is the simplest statistic characterizing the mass
distribution in the universe.  Therefore, the suppression in
clustering due to massive neutrinos is most likely to be observed in
the power spectrum, depicted for several values of neutrino mass in
Fig.~\ref{fig:scott}.
The traditional way of measuring this two-point function is
by analyzing galaxy surveys. Indeed, the most stringent cosmological
constraints on neutrino mass currently come from two galaxy surveys,
the Two Degree Field~\cite{elgaroy} and the Sloan Digital Sky
Survey~\cite{tegmark}.  A cursory glance at the data in
Fig.~\ref{fig:scott} shows
that neutrino masses (actually the constraint is on the sum of all
neutrino masses) greater than several eV are strongly disfavored by
the data. There are two complications to this ``chi-by-eye'' appraisal
though. The first is the concept of {\it bias}: the galaxy
distribution does not necessarily accurately trace the mass
distribution. Bias is thought to be most complicated on scales $k\ga
0.2 h$ Mpc$^{-1}$ so only data on larger scales is typically used to
obtain neutrino constraints. Even with this cut, there is still no
commonly accepted way to treat bias. The considerable scatter in the
upper limits on neutrino mass from galaxy surveys~\cite{nuscatter}
derives mostly from differences in the treatment of bias. The second
complication is that changing other cosmological parameters can
produce similar effects on the power spectrum. These {\it
degeneracies} afflict all measures of the power spectrum and will be
addressed below.

The power spectrum can also be inferred from the Lyman alpha
forest~\cite{croft}. Spectra of distant quasars show absorption at
wavelengths corresponding to the Lyman alpha transition. An absorption
line at wavelength $\lambda$ corresponds to a region of neutral
hydrogen at redshift
$1+z= \lambda/1215\,$\AA.
The clustering of the
lines in the spectra therefore encodes information about the
clustering of the neutral hydrogen and by extension the entire matter
distribution. Again the way to think about the translation between the
observations (flux power spectrum) and the Holy Grail (matter power
spectrum) is that the flux is a biased tracer of the matter. The bias
of the galaxy distribution is thought to be relatively simple on large
scales, but it is difficult to simulate.  The distribution traced by
the forest on the other hand can be simulated quite accurately.
Whereas galaxy formation simulators need to include information about
supernovae, feedback, gas physics, metallicities, and more, the Lyman
alpha forest can be simulated with minor modifications of dark matter
codes~\cite{hui}. Further, the structures probed by the Lyman alpha
forest are at much higher redshift (typically $3-4$) so the clustering
is less developed, more pristine.  Quantitatively, this translates
into the statement that wavenumbers as large as $k\sim 1 h^{-1}$
Mpc$^{-1}$ can be compared confidently with theory, but this is a newer
field of study than galaxy power spectra and therefore less developed.
The systematics
which contaminate the measurements therefore have not yet been fully
explored and accounted for. The data in Fig.~\ref{fig:scott} probably have
optimistic error bars, especially the overall
normalization~\cite{selly}. If indeed the normalization cannot be
pinned down, then data on scales smaller than $1 h$ Mpc$^{-1}$ is
useless as a neutrino probe (recall that the difference in the
spectrum induced by massive neutrinos asymptotes to a constant on
these small scales). Nonetheless, the future in this field appears
bright: the aforementioned galaxy surveys also will take hundreds of
thousands of quasar spectra, so there is hope that the Lyman alpha
forest will produce a robust measurement of the matter power spectrum
at $k \la 1 h$ Mpc$^{-1}$~\cite{mcdon}.
 
Both of the above power spectrum probes have already contributed
useful constraints on neutrino mass. However, there is a third probe,
less developed than the other two, that is potentially even more
powerful and could reach masses as low as $\sqrt{\delta m_{\rm atm}}$:
weak gravitational lensing~\cite{wl}.  Note that at least one neutrino
must have a mass of at least $\sqrt{\delta m_{\rm atm}}$.
Light from distant galaxies is
deflected as it passes through the fluctuating gravitational
potentials along the line of sight. By carefully studying these
deflections, we can glean information about the underlying mass
distribution. The most promising approach is to measure the
ellipticites of many background galaxies. On average the projected 2D
shapes of the galaxies will of course be circular.  Deviations in the
form of non-zero ellipticites carry information about the lensing
field. These deviations are small, typically less than a percent, and
require painstaking observations with careful attention to systematic
problems. The observational status of weak lensing is comparable to
the CMB anisotropy field a decade ago; i.e., it is in its infancy,
just several few years past the initial
detections~\cite{indet}. Still, the community is so excited about weak
lensing because {\it it measures mass directly}. Instead of using
galaxies or absorption lines as mass tracers, the observables are
related to the mass distribution via the simplest tenets of general
relativity. In other words, there is no bias. Current surveys do not
yet have the sky coverage to constrain neutrino mass, but future looks
very promising. There are many ongoing or planned wide-field weak
lensing surveys\,\cite{FutureLens}, and they seem poised to push the
neutrino mass limit down by more than an order of magnitude, close to
$\sqrt{\delta m_{\rm atm}}$.

Changes in other cosmological parameters affect the power spectrum in
ways similar to massive neutrinos. Reducing the total matter density,
for example, suppresses the power spectrum on scales of order $k\sim
0.1 h $ Mpc$^{-1}$ , so supplemental data (typically from the CMB) are
needed to arrive at robust constraints. These degeneracies are often
seen as {\it bad} and as contaminating the cosmological mass
limits. There are two reasons this view should be abandoned, one
technical and the other philosophical. First, it is quite
straightforward to marginalize (integrate) over variations in other
parameters;
current constraints allow for as many as ten other parameters. We are
fortunate that the CMB in particular constrains many of the parameters
to which the matter power spectrum is insensitive. On a deeper level
though, parameter degeneracies connect areas of physics once thought
to be unrelated. As one example, consider the fact that future weak
lensing surveys will be sensitive to the evolution of the power
spectrum at different cosmic epochs. Neutrino masses will affect this
evolution, but so will dark energy~\cite{hutom,kev}. So for example
the spectacular laboratory constraints on neutrino masses anticipated
over the coming decade will break cosmic degeneracies and enable us to
learn about dark energy! The fields of elementary particle physics and
astrophysics are therefore entwined as never before.



\begin{thebibliography}{999}
\bibitem{IceCube} The current status of the ICECUBE project is displayed
 at pheno.physics.wisc.edu/icecube/.
\end{thebibliography}
\end{document}